\documentclass[a4paper,reqno]{amsart}
\usepackage[latin1]{inputenc}

\makeatletter
\RequirePackage{eucal}      % Euler Caligraphics
\RequirePackage{amssymb}    % For double arrows and other

\DeclareMathOperator{\rank}{Rank}
\DeclareMathOperator{\id}{Id}
\DeclareMathOperator{\pr}{pr}
\let\Im\im
\let\ker\Ker

\def\der#1#2{\frac{d#1}{d#2}}
\def\pd#1#2{\frac{\partial#1}{\partial#2}}
\newcommand{\at}[1]{\Big\vert_{#1}}
\newcommand{\tpd}[2]{\partial_{#2}#1}
\newcommand{\tat}[1]{|_{#1}}
\newcommand{\Real}{\mathbb{R}}
\newcommand{\map}[3]{#1\colon#2\rightarrow#3}  
\newcommand{\set}[2]{\left\{\,#1\left.\vphantom{#1#2}\,\right\vert\,#2\,\right\}} %set
\newcommand{\pai}[2]{\langle\,#1\,,#2\,\rangle}  %Canonical pairing
\newcommand{\cinfty}[1]{C^\infty(#1)}

\newcommand{\Sec}[2][]{\operatorname{Sec}\nolimits_{#1}(#2)}
\let\sec\Sec
\let\vf\vectorfields
\let\zerosec\varnothing
\def\hook{\vrule height 0pt depth 0.4pt width 3.5pt
	      \vrule height 5pt depth 0.4pt \kern 3pt}

\newcommand{\spV}{^{\scriptscriptstyle\mathsf{V}}}
\newcommand{\spC}{^{\scriptscriptstyle\mathsf{C}}}
\newcommand{\pb}{{}^\star}

\newcommand{\qquand}{\qquad\text{and}\qquad}
\newcommand{\quand}{\quad\text{and}\quad}

\newcommand{\prol}[2][]{\CMcal{T}^{#1}#2}
\newcommand{\TEE}[1][]{\CMcal{T}^E_{#1}E}

\newcommand{\CJ}[1][E]{\mathcal{C}(J,#1)}
\newcommand{\AJ}[1][E]{\mathcal{A}(J,#1)}
\newcommand{\PJ}[1][E]{\mathcal{P}(J,#1)}

\newcommand{\calo}{{\mathcal{O}}}
\newcommand{\cals}{{\mathcal{S}}}
\newcommand{\calf}{{\mathcal{F}}}

\newcommand{\G}{\boldsymbol{G}}
\renewcommand{\t}{\boldsymbol{t}}
\newcommand{\s}{\boldsymbol{s}}
\newcommand{\bepsilon}{\boldsymbol{\epsilon}}

\newcommand{\g}{\mathfrak{g}}

\newtheoremstyle{slanted}{3pt}{3pt}{\slshape}{}{\scshape}{:}{ }{}
\newtheoremstyle{nonslanted}{3pt}{3pt}{\upshape}{}{\scshape}{:}{ }{}

\theoremstyle{slanted}	
  \newtheorem{definition}{Definition}
  \newtheorem{theorem}{Theorem}
  \newtheorem{proposition}{Proposition}
  \newtheorem{corollary}{Corollary}
  
\theoremstyle{nonslanted}	
  \newtheorem{remarkth}{Remark}
  \newtheorem{exampleth}{Example}

  {\begin{remarkth}}%
  {\hfill$\diamond$\end{remarkth}}

   {\hfill$\triangleleft$\end{exampleth}}

\renewenvironment{proof}
  {\par\normalfont%\footnotesize
   \topsep6\p@\@plus6\p@
   \noindent{\scshape Proof\@addpunct{.}} \ignorespaces}%
  {\qed\endtrivlist\medskip}

\begin{document}

\title[Variational calculus on Lie algebroids]%
      {Variational calculus on Lie algebroids}

\author{Eduardo Mart\'{\i}nez}
\address{Eduardo Mart\'{\i}nez:
Departamento de Matem\'atica Aplicada,
Facultad de Ciencias,
Universidad de Zaragoza, 
50009 Zaragoza, Spain}
\email{emf@unizar.es}
\thanks{Partial financial support from MEC (Spain) grant BFM~2003-02532 is acknowledged}


%\date{March 2002}.
%\date{July 14, 2006}.

\begin{abstract} 
It is shown that the Lagrange's equations for a Lagrangian system on a Lie algebroid are obtained as the equations for the critical points of the action functional defined on a Banach manifold of curves. The theory of Lagrangian reduction and the relation with the method of Lagrange multipliers are also studied.
\end{abstract}

\maketitle

\section{Introduction}
\label{introduction}
The concept of Lie algebroid has proved to be useful in the formulation and analysis of many problems in differential geometry and applied mathematics~\cite{Mackenzie2,CannasWeinstein}.  In the context of Mechanics, a program was proposed by A.\ Weinstein~\cite{Weinstein} in order to develop a theory of  Lagrangian and Hamiltonian systems on Lie algebroids and their discrete analogs on Lie groupoids. In the last years, this program has been actively developed by many authors, and as a result, a powerful mathematical structure is emerging.

One of the main features of the Lie algebroid framework is its inclusive nature. In what respect to Mechanics, under the same formalism one can describe such disparate situations as Lagrangian systems with symmetry, systems evolving on Lie algebras and semidirect products, or systems with holonomic constraints (see~\cite{LSDLA,SLMCLA} for recent reviews) obtaining in such cases Lagrange-Poincaré equations, Poincaré equations, Euler-Poincaré equations or Euler-Lagrange equations for holonomically constrained problems (see~\cite{HoMaRa,CeMaRa,CeMaPeRa,Altafini}, where the theory of Lagrange-Poincaré bundles ---a subclass of transitive Lie algebroids with some additional structure--- is used). One of the advantages of such a unifying formalism is that morphisms establish relations between these apparently different systems, leading to an adequate way to study reduction theory. In addition, by means of an appropriate extension of d'Alembert principle, one can also consider the corresponding versions of such systems when non-holonomic constraints are present~\cite{NHLSLA}. 

While the Lie algebroid approach to Mechanics builds on the geometrical structure of the prolongation of a Lie algebroid~\cite{LMLA} (where one can develop a geometric symplectic treatment of Lagrangian systems parallel to J.\ Klein's formalism~\cite{Cr,Klein}), the origin of Lagrangian Mechanics is the calculus of variations. Integral curves of a standard Lagrangian system are those tangent lifts of curves on the base manifold which are extremal for the action functional defined on a space of paths. 

It is therefore interesting to find a variational description of Lagrange's equations for a Lagrangian system defined on a more general Lie algebroid. The first steps in this direction where already done by A.\ Weinstein in~\cite{Weinstein} in the case of an integrable Lie algebroid (i.e. the Lie algebroid of a Lie groupoid) and by the author in~\cite{Medina,LAGGM}. The purpose of this paper is to analyze the situation for the general case in a solid and rigorous basis. The relevance of having a variational description of such equations is not purely conceptual. It allows to apply the many methods known to solve, simplify, discretize the equations as well as to approximate solutions.

There are many versions of what one calls variational calculus. In full generality, we look for the critical points of a functional defined on a space of functions. To be rigorous enough one has to be precise about the structure of the space of functions where the functional is defined. In general different structures will give different results, and the `same' functional defined on the `same' space but with different topological or differential structure can have or not a solution. In this respect, there are some alternatives for the structure to be required on the space of functions: Banach, Frechet or convenient manifolds are some of the categories used for that, the stronger one being the Banach manifold category.   

We will prove that Lagrange's equations for a Lagrangian system on a Lie algebroid are precisely the equations for the critical points of the action functional defined on the set of admissible curves on a Lie algebroid with fixed base endpoints, in the stronger sense;  that is to say, we will prove that the set of such curves can be endowed with a structure of Banach manifold, that the action functional is continuously differentiable and that the equations for the critical points are precisely Lagrange's equations for the given Lagrangian system as obtained in~\cite{Weinstein,LMLA}. We will also analyse the relation to Lagrange multiplier method and reduction theory following the steps in~\cite{ROCT}.

\subsubsection*{Description of the results and organization of the paper}
Let $\map{\tau}{E}{M}$ be a Lie algebroid with anchor $\map{\rho}{E}{TM}$ and bracket $[\ ,\ ]$. Given a Lagrangian function $L\in\cinfty{E}$ we consider the dynamical system defined locally by the system of differential equations
\[
\begin{aligned}
&\frac{d}{dt}\left(\frac{\partial
  L}{\partial y^\alpha}\right)
+ \frac{\partial L}{\partial y^\gamma}C_{\alpha\beta}^\gamma
y^\beta=\rho_\alpha^i\frac{\partial L}{\partial x^i}\\
&\dot{x}^i=\rho_\alpha^iy^\alpha.
\end{aligned}
\]
The second equation expresses the fact that the curves we have to consider are admissible curves on $E$, also known as $E$-paths, which is the natural concept of path in the category of Lie algebroids. Similarly, there is a natural concept of homotopy of $E$-paths (see~\cite{Rui}). To distinguish from true homotopy we will use the word $E$-homotopy. The base map of an $E$-homotopy is a homotopy with fixed endpoints between the base paths, but in general the converse does not hold.

The set of all $E$-paths defined in the interval $J$ will be denoted $\AJ$. It is clear that $E$-homotopy is an equivalence relation on $\AJ$. It was proved in \cite{Rui} that every $E$-homotopy class is a Banach submanifold of $\AJ$ with codimension equal to the dimension of $E$. The relevant results of~\cite{Rui} are reviewed in section~\ref{homotopy} after some preliminary results in section~\ref{preliminaries}. 

When we consider the action functional $S(a)=\int_{t_0}^{t_1}L(a(t))dt$ defined on a fixed $E$-homotopy class, we will show that the equations for the critical points of $S$ are Lagrange's equations for the Lagrangian $L$ on the Lie algebroid $E$. For the case of integrable Lie algebroids this was already proved in~\cite{Weinstein}.

It may be argued that to restrict to an $E$-homotopy class is not natural. In the standard case $E=TM$, the concept of $E$-homotopy corresponds to the standard notion of homotopy. An $E$-path is just the tangent lift $\dot{\gamma}$ for a given curve $\gamma$ in the base $M$. Two curves $\dot{\gamma}_0$, $\dot{\gamma}_1$ are $TM$-homotopic if and only if there is a homotopy $\phi$ between the base curves with fixed endpoints, the tangent map to $\phi$ being the $TM$-homotopy. In this case every connected component of the (fixed endpoints) path space is a $TM$-homotopy class, and there is no need to select a homotopy class in the variational principle, since they are disconnected sets.

Thus, in the case of a general Lie algebroid, it is natural to endow $\AJ$ with a topology that separates curves in different $E$-homotopy classes. The partition into $E$-homotopy classes defines a foliation on $\AJ$, and hence~\cite{Lang} it defines on the same set $\AJ$ a new Banach manifold structure, which we denote by $\PJ$. This is stated in section~\ref{PathSpace}, as well as some properties of maps induced by morphisms. The action functional $S$ is smooth in such manifold and the variational principle will be stated in section~\ref{variational} in the usual way, that is, by fixing as boundary conditions just and nothing more than the endpoints on the base curve. In particular, the  tangent space to the manifold of admissible paths at an $E$-path (with this adequate differential structure) is spanned by the restriction of complete lifts of (time-dependent) sections of the Lie algebroid to such $E$-path, which are the variations considered in~\cite{Medina,LAGGM}. 

We will also proof that morphisms define mappings between admissible curves and map variations into variations, preserving the variational character of the problem. In particular reduction theory is considered. Finally, in section~\ref{LagrangeMultipliers} we will show that (at least in the integrable case) the problem can be formulated in terms of Lagrange multipliers and that there are not singular points for the constraints. It is also shown an example in which the `heuristic' Lagrange multiplier trick, which is frequently used to solve problems with constraints, is not valid in this case.

\medskip

\paragraph*{\textit{Notation}}
The set of sections of a bundle $\map{\pi}{P}{M}$ will be denoted by $\sec{P}$. When $P=TM$ we will write $\sec{TM}=\vf{M}$. The set of sections of $P$ along a map $\map{f}{N}{M}$ will be denoted by $\sec[f]{P}$. When $P=TM$ we will write $\sec[f]{TM}=\vf{f}$. The notation is as in~\cite{SLMCLA}, except for the canonical involution~\cite{LSDLA} on a Lie algebroid $E$, which will be denoted $\map{\chi_{E}}{\TEE}{\TEE}$. For a curve $\map{a}{\Real}{E}$ and a map $\map{\Phi}{E}{F}$ we will frequently write $\Phi(a)$ instead of $\Phi\circ a$, that is $\Phi(a)(t)=\Phi(a(t))$. This will be particularly useful when we have several curves and the map $\Phi$ takes several  arguments. The projection of the tangent bundle to a manifold $M$ will be denoted $\map{\tau_M}{TM}{M}$.

\section{Preliminaries}
\label{preliminaries}
\subsubsection*{Lie algebroids}
A Lie algebroid structure on a vector bundle $\map{\tau}{E}{M}$ is given by a vector bundle map $\map{\rho}{E}{TM}$ over the identity in $M$, called the anchor, together with a Lie algebra structure on the $\cinfty{M}$-module of sections of $E$ such that the compatibility condition 
$
[\sigma,f\eta]=(\rho(\sigma)f)\eta+f[\sigma,\eta]
$
is satisfied for every $f\in\cinfty{M}$ and every $\sigma,\eta\in\sec{E}$. See~\cite{CannasWeinstein, Mackenzie2} for more information on Lie algebroids.

In what concerns to Mechanics, it is convenient to think of a Lie
algebroid as a generalization of the tangent bundle of $M$. One
regards an element $a$ of $E$ as a generalized velocity, and the
actual velocity $v$ is obtained when applying the anchor to $a$, i.e.,
$v=\rho(a)$. A curve $\map{a}{[t_0,t_1]}{E}$ is said to be
admissible or an $E$-path if $\dot{\gamma}(t)=\rho(a(t))$, where $\gamma(t)=\tau(a(t))$
is the base curve.

A local coordinate system $(x^i)$ in the base manifold $M$ and a local
basis $\{e_\alpha\}$ of sections of $E$, determine a local coordinate system  $(x^i, y^{\alpha})$ on $E$.  The anchor and the bracket are locally determined by the local functions $\rho^i_\alpha$ and $C^\alpha_{\beta\gamma}$ on $M$ given by
\[
\rho (e_{\alpha})=\rho _{\alpha}^{i}\frac{\partial}{\partial x^i}
\qquand
[e_{\alpha}, e_{\beta}]=C_{\alpha\beta}^{\gamma}\ e_{\gamma}.
\]
The functions $\rho^i_\alpha$ and $C^\alpha_{\beta\gamma}$ satisfy some relations due to the compatibility condition and the Jacobi identity which are called the structure equations:
\begin{gather}
 \rho _{\alpha}^{j}\frac{\partial \rho _{\beta}^{i}}{\partial
    x^{j}}-\rho _{\beta}^{j}%
  \frac{\partial \rho _{\alpha}^{i}}{\partial x^{j}}=\rho
  _{\gamma}^{i}C_{\alpha\beta}^{\gamma}~%
  \label{structure.equation.1}
\\
    \rho _{\alpha}^{i}\frac{\partial C_{\beta\gamma}^{\nu}}{\partial x^{i}} +
    \rho _{\beta}^{i}\frac{\partial C_{\gamma\alpha}^{\nu}}{\partial x^{i}} +
    \rho _{\gamma}^{i}\frac{\partial C_{\alpha\beta}^{\nu}}{\partial x^{i}} +
      C_{\beta\gamma}^{\mu}C_{\alpha\mu}^{\nu} +
      C_{\gamma\alpha}^{\mu}C_{\beta\mu}^{\nu} +
      C_{\alpha\beta}^{\mu}C_{\gamma\mu}^{\nu} = 0.
  \label{structure.equation.2}
\end{gather}

\subsubsection*{Cartan calculus} 
The Lie algebroid structure is equivalent to the existence of a exterior differential on $E$, $\map{d}{\Sec{\wedge^kE^*}}{\Sec{\wedge^{k+1}E^*}}$, defined as follows
\begin{align*}
d \omega(\sigma_0,\dots, \sigma_k)&=\sum_{i=0}^{k}
(-1)^i\rho(\sigma_i)(\omega(\sigma_0,\dots,
\widehat{\sigma_i},\dots, \sigma_k))+\\
&+ \sum_{i<j}(-1)^{i+j}\omega([\sigma_i,\sigma_j],\sigma_0,\dots,
\widehat{\sigma_i},\dots,\widehat{\sigma_j},\dots ,\sigma_k),
\end{align*}
for $\omega\in \Sec{\wedge^k E^*}$ and $\sigma_0,\dots ,\sigma_k\in
\Sec{\tau}$. $d$ is a cohomology operator, that is, $d^2=0$. In
particular, if $\map{f}{M}{\Real}$ is a real smooth function then
$df(\sigma)=\rho(\sigma)f,$ for $\sigma\in \Sec{\tau}$. Locally,
\[
dx^i=\rho^i_{\alpha}e^{\alpha}\qquand de^{\gamma}=-\frac{1}{2}
C^{\gamma}_{\alpha\beta} e^{\alpha}\wedge e^{\beta},
\]
where $\{e^{\alpha}\}$ is the dual basis of $\{e_{\alpha}\}$.  The above mentioned structure equations are but the relations $d^2x^i=0$ and $d^2e^\alpha=0$. We may also define the Lie derivative with respect to a section $\sigma$ of $E$ as the operator $\map{d_\sigma}{\Sec{\wedge^k E^*}}{\Sec{\wedge^k E^*}}$ given by $d_\sigma=i_\sigma\circ d+d\circ i_\sigma$.  Along this paper, except otherwise stated, the symbol $d$ stands for the exterior differential on a Lie algebroid.

\subsubsection*{Prolongation}
Given a Lie algebroid $\map{\tau}{E}{M}$ we can consider the vector bundle $\map{\tau_1}{\TEE}{E}$ where the total space is just $\set{(b,v)\in E\times TE}{T\tau(v)=\rho(b)}$, and the projection $\tau_1$ is given by $\tau_1(b,v)=\tau_E(v)$. We will use the redundant notation $(a,b,v)$ for the element $(b,v)$ where $a=\tau_E(v)$, so that $\tau_1$ becomes the projection onto the first factor. The bundle $\TEE$ can be endowed with a structure of Lie algebroid. The anchor $\map{\rho^1}{\TEE}{TE}$ is just the projection onto the third factor $\rho^1(a,b,v)=v$. Local coordinates $(x^i,y^\alpha)$ induce local coordinates $(x^i,y^\alpha,z^\alpha,v^\alpha)$ on $\TEE$, where $z^\alpha$ are the components of $b$ in the basis $\{e_\alpha\}$ and $v^\alpha$ are given by the coordinate expression of $v$, i.e. $b=z^\alpha e_\alpha$ and $v=\rho^i_\alpha z^\alpha\pd{}{x i}+v^\alpha\pd{}{y^\alpha}$. See~\cite{LMLA,SLMCLA} for the definition of the bracket an more details on this Lie algebroid.

Every section $\eta$ of $E$ can be lifted to a section $\eta\spC$ of $\TEE$ given by $\eta\spC(a)=(a,\eta(m),v)$, with $m=\tau(a)$ and where $v\in T_aE$ is the vector that projects to $\rho(\eta(m))$ and satisfies 
\[
v\hat{\theta}=\widehat{d_\eta\theta},
\]
for every section $\theta$ of $E^*$. In this expression, $\hat{\theta}\in\cinfty{E}$ is the linear function associated to the section $\theta\in\sec{E^*}$.
It is clear that the vector field $\rho^1(\eta\spC)\in\vf{E}$ projects to the vector field $\rho(\eta)\in\vf{M}$. Also the section $\eta$ can be lifted vertically to a section $\eta\spV\in\sec{\TEE}$ given by $\eta\spV(a)=(a,0,\eta(m)^v_a)$ where $m=\tau(a)$ and $b^v_a$ denotes the canonical vertical lift of the element $b\in E_m$ to a vertical vector tangent to $E$ at $a$.

The structure of Lie algebroid in $\TEE$ was defined in~\cite{LMLA} in terms of the brackets of vertical an complete lifts
\[
[\eta\spC,\sigma\spC]=[\sigma,\eta]\spC,
\qquad
[\eta\spC,\sigma\spV]=[\sigma,\eta]\spV
\qquand
[\eta\spV,\sigma\spV]=0,
\]
so that we mimic (and hence extend) the properties of complete and vertical lifts in the tangent bundle, which are on the base for the geometric formalism in the calculus of variations.

\subsubsection*{The canonical involution} (See~\cite{LSDLA} for the details.)
There exists a canonical map $\map{\chi_E}{\TEE}{\TEE}$ such that $\chi_E^2=\id$. It is defined by $\chi_E(a,b,v)=(b,a,\bar{v})$, for every $(a,b,v)\in\TEE$, where $\bar{v}\in T_bE$ is the vector which projects to $\rho(a)$ and satisfies 
\[
\bar{v}\hat{\theta}=v\hat{\theta}+d\theta(a,b)
\] 
for every section $\theta$ of $E^*$. 

In particular, for the case of the standard Lie algebroid $E=TM$ we have that $\prol[TM](TM)=TTM$. If we consider a map $\map{\gamma}{\Real^2}{M}$ then $\chi_{TM}$ relates the second partial derivatives of $\gamma$ by
\[
\pd{}{s}\pd{\gamma}{t}=\chi_{TM}\Bigl(\pd{}{t}\pd{\gamma}{s}\Bigr).
\]
In other words, having in mind the calculus of variations, and with the more classical notation $\delta x=\pd{\gamma}{s}$ and $\delta\dot{x}=\pd{}{s}\pd{\gamma}{t}$, we have that $\chi_{TM}$ maps the derivative $\frac{d}{dt}\delta x$ of the variation of the coordinates into the variation of the derivative of the coordinates $\delta \dot{x}$. In the case of a general Lie algebroid, the canonical involution will play a similar role.

In terms of the canonical involution, the complete lift of a section $\eta\in\sec{E}$ is given by
\[
\eta\spC(a)=\chi_E\bigl(\eta(m),a,T_m\eta(\rho(a))\bigr),
\]
with $m=\tau(a)$.

Whenever we have a section defined on an open set we can obtain its complete lift as defined above. Nevertheless if the section is defined only along a curve, we can perform a similar construction with the help of the canonical involution, as it is indicated in the next subsection. 

\subsubsection*{The map $\Xi$}
We will make extensive use of the following map. Given an admissible curve $\map{a}{\Real}{E}$ over $\gamma=\tau\circ a$ we consider the map $\map{\Xi_a}{\sec[\gamma]{E}}{\sec[a]{TE}}$ given by 
\[
\Xi_a(\sigma)=\rho^1(\chi_E(\sigma,a,\dot{\sigma})),
\]
or more explicitly by $\Xi_a(\sigma)(t)=\rho^1(\chi_E(\sigma(t),a(t),\dot{\sigma}(t)))$ for every $t$ in the domain of $a$. In other words, it is determined by $\chi_E(\sigma,a,\dot{\sigma})=(a,\sigma,\Xi_a(\sigma))$. From the definition it is easy to prove the following property
\[
\Xi_a(f\sigma)=f\Xi_a(\sigma)+\dot{f}\sigma^v_a,
\]
for every function $f\in\cinfty{\Real}$.

Complete lifts can be obtained in terms of the above map. If $a$ is an admissible curve over $\gamma$, then $\eta\spC(a(t))=\chi_E\bigl(\eta(\gamma(t)),a(t),\der{}{t}(\eta(\gamma(t)))\bigr)$ in other words,
\[
\rho^1(\eta\spC)\circ a=\Xi_a(\eta\circ \gamma). 
\]
The above relation can serve to define the complete lift of a time-dependent section $\eta(t,m)$ by taking $t\mapsto \eta(t,\gamma(t))$ instead of $\eta\circ\gamma$ above. Nevertheless, in the next section we will follow another approach.

We will see later on that for our problem, admissible infinitesimal variations of an admissible curve $a$ are of the form $\Xi_a(\sigma)$.

\subsubsection*{Local expressions}
In local coordinates, if $\eta=\eta^\alpha e_\alpha$ is a local section of $E$ then the vector field associated to its complete lift has the local expression
\[
\rho^1(\eta\spC)=\rho^i_\alpha\eta^\alpha+
\Bigl(\rho^i_\beta y^\beta\pd{\eta^\alpha}{x^i} +C^\alpha_{\beta\gamma}y^\beta\eta^\gamma\Bigr)\pd{}{y^\alpha}.
\]
The canonical involution is locally given by 
\[
\chi_E(x^i,y^\alpha,z^\alpha,v^\alpha)=(x^i,z^\alpha,y^\alpha,v^\alpha+C^\alpha_{\beta\gamma}z^\beta y^\gamma).
\]
Finally the expression of the map $\Xi_a$ is 
\[
\Xi_a(\sigma)(t)=\rho^i_\alpha(\gamma(t))\sigma^\alpha(t)\pd{}{x^i}\at{a(t)}
+\Bigl(\dot{\sigma}^\alpha(t)+C^\alpha_{\beta\gamma}(\gamma(t))a^\beta(t)\sigma^\gamma(t)\Bigr)\pd{}{y^\alpha}\at{a(t)}.
\]
where $a$ and $\sigma$ have the local expression $a(t)=(\gamma^i(t),a^\alpha(t))$ and $\sigma(t)=(\gamma^i(t),\sigma^\alpha(t))$.

\subsubsection*{Lagrangian systems}
Given a function $L\in\cinfty{E}$, we define a dynamical system on $E$ by means of a system of differential equations (see~\cite{Weinstein}) which in local coordinates reads
\begin{equation}
\label{Lagrange}
\begin{aligned}
&\frac{d}{dt}\left(\frac{\partial
  L}{\partial y^\alpha}\right)
+ \frac{\partial L}{\partial y^\gamma}C_{\alpha\beta}^\gamma
y^\beta=\rho_\alpha^i\frac{\partial L}{\partial x^i}\\
&\dot{x}^i=\rho_\alpha^iy^\alpha.
\end{aligned}
\end{equation} 
The second equation express is just the condition of admissibility for a curve. Therefore we have to look for admissible curves satisfying the first equation. 

In geometric terms the above dynamical system can be obtained as follows (see~\cite{LMLA,SLMCLA} for intrinsic definitions, detailed proofs and an alternative symplectic setup). Associated to $L$ there is a section  $\theta_L$ of $(\TEE)^*$, such that $\pai{\theta_L}{\eta\spC}=d_{\eta\spV}L$ and $\pai{\theta_L}{\eta\spV}=0$. A solution of Lagrange's equations is an admissible curve $\map{a}{\Real}{E}$ which satisfies $\delta L(\dot{a}(t))=0$, where 
\[
\pai{\delta L(\dot{a}(t))}{\eta(\gamma(t))}
=\bigl(d_{\eta\spC}L\bigr)(a(t)) -\frac{d}{dt}\bigl(\pai{\theta_L}{\eta\spC}(a(t))\bigr)
\]
for every $\eta\in\sec[\gamma]{E}$ and where $\gamma=\tau\circ a$ is the curve on the base. It is easy to see that the above expression depends linearly on $\eta$, and the vanishing of $\delta L$ is equivalent to the above system of differential equations~\eqref{Lagrange}.

In what follows, we will identify $\theta_L$ with the Legendre map, so that by the expression 
$\pai{\theta_L}{\eta}$ we mean $\pai{\theta_L}{\tilde{\eta}}$ for any section $\tilde{\eta}$ of $\TEE$ projecting to $\eta$. In other words $\pai{\theta_L}{\eta}=d_{\eta\spV}L$ and in coordinates $\theta_L(x,y)=\pd{L}{y^\alpha} e^\alpha$.

\subsubsection*{Morphisms}
Given a second Lie algebroid $\map{\tau'}{E'}{M'}$, a vector bundle map $\map{\Phi}{E}{E'}$ over $\map{\varphi}{M}{M'}$ is said to be admissible if it maps admissible curves in $E$ into admissible curves in $E'$, or equivalently if $\rho'\circ\Phi=T\varphi\circ\rho$. The map $\Phi$ is said to be a morphism of Lie algebroids if $\Phi\pb d\theta=d\Phi\pb\theta$ for every $p$-form $\theta\in\Sec{\wedge^pE^*}$. Every morphism is an admissible map.

In coordinates, a vector bundle map $\Phi(x,y)=(\varphi^i(x),\Phi^\alpha_\beta(x)y^\beta)$ is admissible if and only if 
\begin{equation}
\rho'{}^i_\alpha\pd{\varphi^k}{x^i}=\rho^k_\beta\Phi^\beta_\alpha.
\end{equation}
Moreover, such a map is a morphism if in addition to the above equation it satisfies
\begin{equation}
\rho^i_\nu\pd{\Phi^\alpha_\mu}{x^i}-\rho^i_\mu\pd{\Phi^\alpha_\nu}{x^i}-
C'{}^\alpha_{\beta\gamma}\Phi^\beta_\mu\Phi^\gamma_\nu=0.
\end{equation}

Important examples of Lie algebroid morphism are the following. If $\eta$ is a section of $E$ then the flow $\Phi_s$ of the vector field $\rho^1(\eta\spC)\in\vf{E}$ projects to the flow $\varphi_s$ of the vector field $\rho(\eta)\in\vf{M}$. For every fixed $s$, the map $\Phi_s$ is a vector bundle map which is a morphism of Lie algebroids over $\varphi_s$. The pair $(\Phi_s,\varphi_s)$ is said to be the flow of the section $\eta\in\sec{E}$, and we have that 
\[
d_\eta\theta=\frac{d}{ds}\Phi\pb_s\theta\at{s=0},
\]
for every tensor field $\theta$ over $E$.

Given an admissible map $\map{\Phi}{E}{E'}$ we can define a map $\map{\prol[\Phi]{\Phi}}{\TEE}{\prol[E']{E'}}$ by means of $\prol[\Phi]{\Phi}(a,b,v)=(\Phi(a),\Phi(b),T_a\Phi(v))$ for every $(a,b,v)\in\TEE$.

\begin{proposition}
\label{morphism}
Let $\map{\Phi}{E}{E'}$ be an admissible map. The following conditions are equivalent
\begin{enumerate}
\item $\Phi$ is a Lie algebroid morphism.
\item $\prol[\Phi]{\Phi}$ is a Lie algebroid morphism.
\item $\prol[\Phi]{\Phi}\circ\chi_E=\chi_{E'}\circ\prol[\Phi]{\Phi}$.
\item $T{\Phi}\circ\Xi_a(\sigma)=\Xi_{\Phi\circ a}(\Phi\circ\sigma)$ for every $E$-path $a$ and every section $\sigma$ along the base path $\tau\circ a$.
\end{enumerate}
\end{proposition}
\begin{proof}
The equivalence between (1) and (2) was proved in~\cite{CFTLAMF}. To prove the equivalence between (1) and (3) we use the definition of the canonical involution. For $(a,b,v)\in\TEE$ we have, on one hand
\[
\prol[\Phi]{\Phi}(\chi_E(a,b,v))
=\prol[\Phi]{\Phi}(b,a,\bar{v})
=(\Phi(b),\Phi(a),T\Phi(\bar{v})),
\]
and on the other 
\[
\chi_{E'}(\prol[\Phi]{\Phi}(a,b,v))
=\chi_{E'}(\Phi(a),\Phi(b),T\Phi(v))
=(\Phi(b),\Phi(a),\overline{T\Phi(v)}).
\]
Therefore we have to prove the equivalence of the morphism condition with the condition $\overline{T\Phi(v)}=T\Phi(\bar{v})$ for every $v$ as above. For every $\theta\in\sec{E^*}$ we have 
\begin{align*}
[\overline{T\Phi(v)}-T\Phi(\bar{v})]\hat{\theta}
&=T\Phi(v)\hat{\theta}+d\theta(\Phi(a),\Phi(b))
 -\bar{v}\,\widehat{\Phi\pb\theta}\\
&=v\,\widehat{\Phi\pb\theta}+(\Phi\pb d\theta)(a,b)
 -v\,\widehat{\Phi\pb\theta}-d(\Phi\pb\theta)(a,b)\\
&=(\Phi\pb d\theta)(a,b)-d(\Phi\pb\theta)(a,b).
\end{align*}
Since $\Phi$ is admissible, the vanishing of the left-hand side is equivalent to $\Phi$ being a morphism.

Condition (3) implies (4), by just evaluating (3) on elements of the form $(\sigma,a,\dot{\sigma})$,
\begin{align*}
\prol[\Phi]{\Phi}\circ\Xi_a(\sigma)
&=\prol[\Phi]{\Phi}\circ\chi_E(\sigma,a,\dot{\sigma})\\
&=\chi_{E'}\circ\prol[\Phi]{\Phi}(\sigma,a,\dot{\sigma})\\
&=\chi_{E'}(\Phi\circ \sigma,\Phi\circ a,T\Phi\circ\dot{\sigma})\\
&=\chi_{E'}(\Phi\circ \sigma,\Phi\circ a,\frac{d}{dt}(\Phi\circ\sigma))\\
&=\Xi_{\Phi\circ a}(\Phi\circ\sigma).
\end{align*}
Finally the difference $D=T{\Phi}\circ\Xi_a(\sigma)-\Xi_{\Phi\circ a}(\Phi\circ\sigma)$ is a vertical vector and evaluating it on the coordinates $y^\alpha$ we easily get 
\[
D(t)\cdot y^\alpha
=\left(
\rho^i_\nu\pd{\Phi^\alpha_\mu}{x^i}-\rho^i_\mu\pd{\Phi^\alpha_\nu}{x^i}-
C'{}^\alpha_{\beta\gamma}\Phi^\beta_\mu\Phi^\gamma_\nu
\right)a^\mu\sigma^\nu.
\]
This implies condition (1) and  ends the proof.

A more intrinsic proof of the last equivalence can be obtained by proving first that  $D(t)\cdot\hat{\theta}=(\sigma\hook(d\Phi\pb\theta-\Phi\pb d\theta))^{\wedge}$ for every section $\theta$ of $E^*$.
\end{proof}
For our proposals, the importance of the above proposition is that item (4) ensures that morphisms map infinitesimal variations into infinitesimal variations.
 
\section{Homotopy of $E$-paths}
\label{homotopy}
The content of this section is mainly a recompilation of some of the results in~\cite{Rui}, slightly reformulated in a more appropriate way for our proposals. It may be summarized as follows: In the set of $E$-paths we can define a equivalence relation known as $E$-homotopy. The space of $E$-paths is a Banach manifold, and every $E$-homotopy class is a smooth Banach submanifold. The partition into $E$-homotopy classes defines a foliation. The tangent space to that foliation at an $E$-path $a$ is the image by the map $\Xi_a$ of the set of sections along $a$ which vanish at the endpoints.

\subsection{$E$-homotopy defined}
As we said above, a curve $a$ on a Lie algebroid $E$ is said to be admissible or an $E$-path if it satisfies $\rho\circ a=\dot{\gamma}$, where $\gamma$ is the base curve $\gamma=\tau\circ a$. Alternatively, an $E$-path can be considered as a morphism of Lie algebroids $\map{a\,dt}{T\Real}{E}$.  In the category of Lie algebroids there is a natural concept of homotopy of $E$-paths. To distinguish it from true homotopy we will refer to it as $E$-homotopy.

Let $I=[0,1]$ and $J=[t_0,t_1]$, and denote the coordinates in $\Real^2$ by $(s,t)$. 
Given a vector bundle map $\map{\Phi}{T\Real^2}{E}$, denote $a(s,t)=\Phi(\tpd{}{t}\tat{(s,t)})$ and $b(s,t)=\Phi(\tpd{}{s}\tat{(s,t)})$, so that we can write $\Phi=adt+bds$.

\begin{definition}
Two $E$-paths $a_0$ and $a_1$ are said to be $E$-homotopic if there exists a morphism  of Lie algebroids $\map{\Phi}{TI\times TJ}{E}$, $\Phi=a dt+bds$, such that 
\begin{align*}
&a(0,t)=a_0(t)
&&b(s,t_0)=0\\
&a(1,t)=a_1(t)
&&b(s,t_1)=0.
\end{align*}
We will say that $\Phi$ is an $E$-homotopy from the $E$-path $a_0$ to the $E$-path $a_1$.
\end{definition}

From the definition it follows that the base map is a homotopy in $M$ from the base path $\tau\circ a_0$ to the base path $\tau\circ a_1$ with fixed endpoints. Moreover, the admissibility conditions for the map $\Phi$ are just the conditions for the curves $t\mapsto a(s,t)$ and $s\mapsto b(s,t)$ to be admissible curves. Finally 

\begin{proposition}
An admissible map $\Phi=adt+bds$ is a morphism if and only if 
\begin{equation}
\chi_E\left(b,a,\pd{b}{t}\right)=\left(a,b,\pd{a}{s}\right).
\end{equation}
\end{proposition}
\begin{proof}
It can be easily deduced from the results in~\cite{LSDLA}, but a coordinate calculation readily shows that the above condition is equivalent to 
\[
\pd{b^\gamma}{t}-\pd{a^\gamma}{s}+C^\gamma_{\alpha\beta}a^\alpha b^\beta=0.
\]
On the other hand, this is just the condition $\Phi\pb d\theta-d\Phi\pb\theta=0$, for $\theta$ the elements of the basis of sections of $E^*$ associated to a linear coordinate system on $E$.
\end{proof}

It follows that if $\Phi=adt+bds$ is a morphism, then the vector tangent to the variation curve $s\mapsto a(s,t)$ is 
\[
\pd{a}{s}(s,t)=\Xi_{a_s}(b_s)
\]
where we have written $a_s(t)=a(s,t)$ and $b_s(t)=b(s,t)$. In particular, at $t=0$ if we write $\sigma(t)=b(0,t)$ then 
$
\pd{a_s}{s}\at{s=0}=\Xi_{a_0}(\sigma)
$.

\subsection{Construction of $E$-homotopies}
From a section of $E$ and an $E$-path we can construct a morphism from $T\Real^2$ to $E$ as indicated in the following proposition. We first recall that the flow of a section $\eta$ is a morphism of Lie algebroids $(\Phi_s,\varphi_s)$, where $\Phi_s$ is the flow of the vector field $\rho^1(\eta\spC)$ and $\varphi_s$ is the flow of the vector field $\rho(\sigma)$.

\begin{proposition}
\label{morphism.construction}
Let $a_0$ be an $E$-path, with base path $\gamma_0$, and let $\eta$ be a section of~$E$. Denote by $(\Phi_s,\varphi_s)$ the flow of the section $\eta$ and define 
\[
a(s,t)=\Phi_s(a_0(t))
\qquad
\gamma(s,t)=\varphi_s(\gamma_0(t))
\qquand
b(s,t)=\eta(\gamma(s,t)).
\]
Then $\xi=a(s,t)dt+b(s,t)ds$ is a morphism from $T\Real^2$ to $E$ over $\gamma$. 
\end{proposition}
\begin{proof}
Since $\Phi_s$ projects to $\varphi_s$, it is clear that $\xi$ is a vector bundle map over $\gamma$. We first prove that $\xi$ is admissible. It is clear that $t\mapsto a(s,t)$ is admissible since $a_0$ is admissible and $\Phi_s$ is a morphism. For the curve $s\mapsto b(s,t)$ we have 
\[
\pd{\gamma}{s}(s,t)
=\pd{}{s}\varphi_s(\gamma_0(t))
=\rho(\eta)(\varphi_s(\gamma_0(t)))
=\rho(\eta(\gamma(s,t)))
=\rho(b(s,t)),
\]
where we have used that $\varphi_s$ is the flow of $\rho(\eta)$.
Finally we prove that $\chi_E(b,a,\pd{b}{t})=(a,b,\pd{a}{s})$. On one hand,
$\pd{b}{t}(s,t)=T\eta(\pd{\gamma}{t})=T\eta(\rho(a(s,t)))$, from where we get 
\[
\chi_E\Bigl(b,a,\pd{b}{t}\Bigr)=\chi_E(\eta(\gamma),a,T\eta(\rho(a)))=\eta\spC(a).
\]
On the other hand
\[
\pd{a}{s}(s,t)
=\pd{}{s}(\Phi_s(a_0(t)))
=\rho^1(\eta\spC)(\Phi_s(a_0(t)))
=\rho^1(\eta\spC)(a(s,t)),
\]
so that $\bigl(a,b,\pd{a}{s}\bigr)=(\eta(\gamma),a,\rho^1(\eta\spC)(a))=\eta\spC(a)$, and both expressions coincide.
\end{proof}

\begin{corollary}
Let $m_0,m_1$ be two points in $M$ and let $\eta$ be a section of $E$ with compact support such that $\eta(m_0)=\eta(m_1)=0$. Let $a_0$ be a curve such that its base curve connects the point $m_0$ with $m_1$. Then the map $\xi$, constructed as in proposition~\ref{morphism.construction}, is an $E$-homotopy from $a_0$ to $a_1=\Phi_1\circ a_0$.
\end{corollary}
\begin{proof}
The condition of compact support implies that the flow of $\eta$ is globally defined, so that $\Phi_1$ is defined. Obviously $a(0,t)=a_0$ and $a(1,t)=a_1$. Since $\rho(\eta)$ vanishes at $m_0$ we have that $\varphi_s(m_0)=m_0$ from where 
\[
b(s,t_0)=\eta(\gamma(s,t_0))=\eta(\varphi_s(\gamma(t_0)))=\eta(\varphi_s(m_0))=\eta(m_0)=0.
\]
A similar argument shows that $b(s,t_1)=0$.
\end{proof}

%% Time dependent

\subsubsection*{Extension to time-dependent sections}
We need to extend the above result to time dependent sections and flows (to ensure the existence of a solution $\eta$ for an equation such as $\sigma(t)=\eta(t,\gamma(t))$ for a section $\sigma$ along $\gamma$, which may not have solution for $\eta$ time-independent). The best way to treat them is to move to the time-dependent setting. Given the Lie algebroid $\map{\tau}{E}{M}$ we consider the direct product Lie algebroid of $T\Real$ with $E$, that is, the Lie algebroid $\bar{\tau}\colon\bar{E}=T\Real\times E\to\bar{M} =\Real\times M$ with anchor $\bar{\rho}(\lambda\tpd{}{t}+a)=\lambda\tpd{}{t}+\rho(a)$ and bracket determined by the bracket of biprojectable sections $[\alpha+\eta,\beta+\zeta]=[\alpha,\beta]_{T\Real}+[\eta,\zeta]_E$. The projections $\pr_1$ and $\pr_2$  onto the factors are morphisms of Lie algebroids.

Every curve $\gamma$ on $M$ can be lifted to a curve $\bar{\gamma}$ on $\bar{M}$ by $\bar{\gamma}(t)=(t,\gamma(t))$. Every curve $a$ in $E$ can be lifted to a curve $\bar{a}$ in $\bar{E}$ by $\bar{a}(t)=\tpd{}{t}\tat{t}+a(t)$. With this definitions, it is obvious that $\bar{a}$ is admissible in $\bar{E}$ if and only if $a$ is admissible in~$E$. A time-dependent section $\eta$ of $E$ can be lifted to a section of $\bar{E}$ by $\bar{\eta}(t,m)=\eta(t,m)=0\cdot\tpd{}{t}\tat{t}+\eta(t,m)$. We have that $\pr_2\circ\bar{\gamma}=\gamma$, $\pr_2\circ\bar{a}=a$ and $\pr_2\circ\bar{\eta}=\eta$. 

With all this considerations at hand we can extend the previous result for time dependent sections as follows:

Let $a_0$ be an $E$-path, with base path $\gamma_0$, and let $\eta$ be a time-dependent section of $E$. Consider the time dependent lifting $\bar{a}_0$ of $a_0$ and $\bar{\eta}$ of $\eta$ as above. Let $\map{\bar{\xi}}{T\Real^2}{\bar{E}}$ the morphism constructed as in proposition~\ref{morphism.construction}. Then $\xi=\pr_2\circ\bar{\xi}$ is a morphism from $T\Real^2$ to $E$. Explicitly, if $(\Phi_s,\varphi_s)$ is the flow of the section $\bar{\eta}$, then $\bar{\xi}=\Phi_s(\bar{a}_0(t))dt+\bar{\eta}(\bar{\gamma}(s,t))ds$, with $\bar{\gamma}(s,t)=\varphi_s(t,\gamma_0(t))$. If we set  $\gamma(s,t)=\pr_2(\bar{\gamma}(s,t))$, then
\[
\xi=\pr_2\bigl(\Phi_s(\tpd{}{t}\tat{t}+a_0(t))\bigr) dt+\eta(t,\gamma(s,t))ds,
\]
in other words $a(s,t)=\pr_2\bigl(\Phi_s(\tpd{}{t}\tat{t}+a_0(t))\bigr)$ and $b(s,t)=\eta(t,\gamma(s,t))$.

Moreover, if $\eta$ has compact support (so that the flow of $\bar{\eta}$ is globally defined) and $\eta(t_0,m_0)=\eta(t_1,m_1)=0$, where $m_0=\gamma_0(t_0)$ and $m_1=\gamma_0(t_1)$, then $\xi$ is a homotopy from $a_0$ to the curve $a(1,-)$. 

Since $\xi$ is a morphism it follows that $(a,b,\pd{a}{s})=\chi_E(b,a,\pd{b}{t})$, so that 
\[
\pd{a}{s}(s,t)=\rho^1\left(\chi_E\Bigl(\eta(t,\gamma(s,t)),a(s,t),\pd{\eta}{t}(s,t)+T\eta_t\bigl(\rho(a(s,t))\bigr)\Bigr)\right).
\]
which we can simply write in the form
\[
\pd{a}{s}=\rho^1\Bigl(\chi_E\Bigl(\eta(t,\gamma),a,\pd{\eta}{t}+T\eta_t\bigl(\rho(a)\bigr)\Bigr)\Bigr).
\]
In other words $\pd{a}{s}(s,t)=\Xi_a(\eta\circ\bar{\gamma})$ and $\pd{a}{s}(0,t)=\Xi_a(\sigma)(t)$ with $\sigma(t)=\eta(t,\gamma(t))$. Therefore, the vectors of the form $\Xi_a(\sigma)$ are tangent to (curves contained in) an $E$-homotopy class.

\subsection{Differentiable structure}
$E$-homotopy, being an equivalence relation, defines a partition of the space of $E$-paths into disjoint sets. We will now proof that every $E$-homotopy class is a smooth Banach manifold and that such partition is a foliation.

For a vector bundle $\map{\pi}{F}{M}$ we consider the set $\CJ[F]$ of all $C^1$ curves $\map{a}{J}{F}$ such that the base path $\gamma=\tau\circ a$ is $C^2$. It is well known that $\CJ[F]$ is a Banach manifold (see~\cite{MTA,PiTa}). In particular we will consider the case of curves on the tangent bundle, $F=TM$, and the case of curves on our Lie algebroid, $F=E$.  The set of $E$-paths
\[
\AJ=\set{\map{a}{J}{E}}{\rho\circ a=\frac{d}{dt}(\tau\circ a)}
\]
is a subset of $\CJ$. 

We will prove first that $\AJ$ is a Banach manifold. 
For that we consider the map $\map{G}{\CJ}{\CJ[TM]}$ given by $G(a)=\frac{d}{dt}(\tau\circ a)-\rho\circ a$. Notice that $G(a)$ is a curve over $\tau\circ a$. If $\calo\subset\CJ[TM]$ is the set of curves in $TM$ contained in the zero section, then it is clear that $\AJ=G^{-1}(\calo)$. 

The tangent map to $G$ at a point $a\in\CJ$ is given by 
\[
T_aG(V)=\chi_{TM}\frac{d}{dt}(T\tau\circ V)-T\rho\circ V.
\]
for $V\in T_a\CJ$. Indeed, take a curve $a_s$ in $\CJ$ such that $a_0=a$ and denote $V=\pd{a_s}{s}\at{s=0}$. Then 
\begin{align*}
T_aG(V)(t)
&=\pd{}{s}\at{s=0}G(a_s)(t)\\
&=\pd{}{s}\pd{}{t}(\tau(a_s(t)))\at{s=0}-\pd{}{s}\rho(a_s(t))\at{s=0}\\
&=\chi_{TM}\pd{}{t}\pd{}{s}(\tau(a_s(t)))\at{s=0}-T\rho(V(t))\\
&=\chi_{TM}\frac{d}{dt}(T\tau(V(t)))-T\rho(V(t))
\end{align*}
that proves the result.

For the proof of the next proposition, we recall that a Banach subspace $\map{i}{F}{E}$ splits if there exists an isomorphism $\map{\alpha}{E}{F_1\times F_2}$, where $F_1, F_2$ are Banach spaces, such that $\alpha\circ i$ induces an isomorphism from $F$ to $F_1\times\{0\}$.

\begin{proposition}
The map $G$ is transversal to the submanifold $\mathcal{O}$. The set $\AJ$ is a Banach submanifold of $\CJ$.
\end{proposition}
\begin{proof}
We prove that $G$ is transversal to $\mathcal{O}$ at any point $a\in\AJ$. Indeed, if $\gamma=\tau\circ a$ and $\map{\zerosec}{M}{TM}$ denotes the zero section of the tangent bundle $\map{\tau_M}{TM}{M}$, then $G(a)=\zerosec\circ\gamma$, and the transversality condition means that for every $X\in\vf{\zerosec\circ\gamma}$ there exists $U$ tangent to the zero section and $Z\in T_a\CJ$ such that $T_aG(Z)+U=X$. If $U$ is tangent to the zero section then it is of the form $U=T\zerosec\circ R$ for some $R\in\vf{\gamma}$. Therefore $T\tau_M\circ U=R$ and since $T_aG(Z)$ is vertical, we get that $R=T\tau_M\circ X$. Therefore, the transversality condition means that given $X$, there exists $Z$ such that 
\[
\chi_{TM}\frac{d}{dt}(T\tau\circ V))-T\rho\circ V=(\id-T\zerosec\circ T\tau_M)\circ X.
\]
This is a linear non homogeneous ordinary differential equation for $T\tau\circ V$ which has always a global solution. In coordinates, if $V=W^i\pd{}{x^i}+Z^\alpha\pd{}{y^\alpha}$ and 
$X=R^i\pd{}{x^i}+X^i\pd{}{v^i}$ then $X-T\zerosec\circ T\tau_M\circ X=X^i\pd{}{v^i}$ and the above equation reads
\[
\dot{W}^i-W^j\pd{\rho^i_\alpha}{x^j}a^\alpha=Z^\alpha\rho^i_\alpha+X^i.
\]

On the other hand, $(T_aG)^{-1}(T\mathcal{O})$ splits. Indeed, we have that  
\[
(T_aG)^{-1}(T\mathcal{O})=\set{b^v_a\in\vf{a}}{b\in\sec[\gamma]{E}\text {such that }\rho(b)=0}
=\sec{\ker(\rho)^v_a}.
\]
Since $a$ is admissible,  $\rho$ has constant rank along $a$ and therefore we can find a subbundle $\mathcal{Z}$ of $a^*(TE)$ such that $a^*(TE)=\ker(\rho)^v_a\oplus\mathcal{Z}$. Thus $T_a\CJ=\sec[a]{E}=\sec{\ker(\rho)^v_a}\oplus\sec{\mathcal{Z}}=(T_aG)^{-1}(T\mathcal{O})\oplus\sec{\mathcal{Z}}$.

Therefore $G$ is transversal to the submanifold of paths contained in the zero section from where it follows that $\AJ$ is a Banach submanifold of $\CJ$ and its tangent space at $a\in\AJ$ is the kernel of $T_aG$.
\end{proof}

\medskip

Vectors of the form $\Xi_a(\sigma)$ can be easily seen to be elements of $T_a\AJ=\ker T_aG$, but not every element in $\ker T_aG$ is of that form. Obviously $Z=\Xi_a(\sigma)$ satisfies $T\tau\circ Z(t)\in\Im\rho_{\gamma(t)}$ for every $t\in J$. Conversely,

\begin{proposition}
Let $Z\in T_a\AJ$ for $a\in\AJ$ and put $\gamma=\tau\circ a$. There exists $\sigma\in\Sigma_\gamma$ such that $Z=\Xi_a(\sigma)$ if and only if there exists $t_2\in J$ such that $T\tau\circ Z(t_2)\in\Im\rho_{\gamma(t_2)}$. Moreover, if we fix $b\in E_{t_2}$ such that $T\tau\circ Z(t_2)=\rho(b)$, then there exists a unique $\sigma$ such that $Z=\Xi_a(\sigma)$ and $\sigma(t_2)=b$.
\end{proposition}
\begin{proof}
We will work in local coordinates (see~\cite{Rui} pag.~605 for a more intrinsic proof using an auxiliary connection). 

Let $Z\in\ker(T_aG)$ and denote $W=T\tau\circ Z$, so that $W(t_2)=\rho(b)$ for some $b\in E$. If the coordinate expression of $Z$ is $Z=W^i\pd{}{x^i}+Z^\alpha\pd{}{y^\alpha}$  then $W=W^i\pd{}{x^i}$ and we have $W^i(t_2)=\rho^i_\alpha b^\alpha$. The condition $T_aG(Z)=0$ reads in coordinates 
\[
\dot{W}^i=W^j\pd{\rho^i_\alpha}{x^j}a^\alpha+Z^\alpha\rho^i_\alpha.
\]
We consider the auxiliary initial value problem for a linear ordinary differential equation given by
\[
\dot{\sigma}^\alpha+a^\gamma C^\alpha_{\gamma\beta}\sigma^\beta=Z^\alpha
\qquad\qquad
\sigma^\alpha(t_2)=b^\alpha.
\]
This equation has a global solution $\sigma(t)$ defined on the interval $J$. We will prove that $W=\rho\circ\sigma$. The difference  $D(t)=W(t)-\rho_{\gamma(t)}(\sigma(t))$ satisfies the homogeneous linear differential equation
\[
\dot{D}^i=\left(a^\alpha\pd{\rho^i_\alpha}{x^j}\right)D^j.
\]
Indeed,
\begin{align*}
\dot{D}^i
&=\dot{W}^i-\pd{\rho^i_\alpha}{x^j}\dot{x}^j\sigma^\alpha -\rho^i_\alpha\dot{\sigma}^\alpha\\
&=\left(W^j\pd{\rho^i_\alpha}{x^j}a^\alpha+Z^\alpha\rho^i_\alpha\right)
 - \pd{\rho^i_\alpha}{x^j}\rho^j_\beta a^\beta\sigma^\alpha -\rho^i_\alpha\left(Z^\alpha-a^\gamma C^\alpha_{\gamma\beta}\right)\\
%&=\pd{\rho^i_\alpha}{x^j}a^\alpha(W^j-\rho^j_\beta\sigma^\beta)
%+\left(\rho^j_\beta\pd{\rho^i_\alpha}{x^j}-\rho^j_\alpha\pd{\rho^i_\beta}{x^j}
%+\rho^i_\gamma C^\gamma_{\alpha\beta}\right)a^\alpha\sigma^\beta\\
&=\pd{\rho^i_\alpha}{x^j}a^\alpha(W^j-\rho^j_\beta\sigma^\beta)\\
&=\pd{\rho^i_\alpha}{x^j}a^\alpha D^j,
\end{align*}
where we have used the first structure equation~\eqref{structure.equation.1}. Since moreover $D(t_2)=W(t_2)-\rho(\sigma(t_2))=0$ we deduce that $D(t)=0$ for all $t\in J$, and hence $W=\rho\circ\sigma$.

As a consequence, the coordinate expression of $Z$ is 
\[
Z=\rho^i_\alpha\sigma^\alpha\pd{}{x^i}+(\dot{\sigma}^\alpha+a^\gamma C^\alpha_{\gamma\beta}\sigma^\beta)\pd{}{y^\alpha},
\]
which is but the local expression of $\Xi_a(\sigma)$, for $\sigma=\sigma^\alpha(t) e_\alpha(\gamma(t))$.

From the construction of $\sigma$ above, we have that $\sigma(t_2)=b$. We now prove that it is unique. If there are two sections $\sigma$ and $\sigma'$ such that $\Xi_a(\sigma)=\Xi_a(\sigma')$ and $\sigma(t_2)=b=\sigma'(t_2)$ then the difference $\lambda=\sigma-\sigma'$ satisfies $\Xi_a(\lambda)=0$ and $\lambda(t_2)=0$. This implies that the components of $\lambda$ are the solution of the initial value problem 
\[
\dot{\lambda}^\alpha+a^\gamma C^\alpha_{\gamma\beta}\lambda^\beta=0
\qquad\qquad
\lambda^\alpha(t_2)=0,
\]
so that $\lambda$ vanishes and hence $\sigma=\sigma'$.
\end{proof}

In other words, vector fields in $T_a\AJ$ are either tangent to a leaf of the Lie algebroid or transversal to all the leaves. We will frequently use the following particular case.

\begin{corollary}
\label{Xi.injective}
Let $Z\in T_a\AJ$ for $a\in\AJ$ such that $T\tau\circ Z(t_0)=0$ then there exists a unique $\sigma\in\sec[\gamma]{E}$ such that $Z=\Xi_a(\sigma)$ and $\sigma(t_0)=0$.
\end{corollary}

\bigskip

If an $E$-homotopy class is to be a manifold, then we have seen that the tangent space contains the vectors of the form $\Xi_a(\sigma)=\rho^1(\chi_E(\sigma,a,\dot{\sigma}))$ (tangent to the curve $s\mapsto a_s)$. We will prove that this vectors define an integrable distribution whose leaves are precisely the homotopy classes, and as a consequence, such kind of vectors span the whole tangent space to the given homotopy class.

\begin{definition}
For $a\in\AJ$, denote by $\gamma$ the base curve and define the vector space
\[
\Sigma_\gamma=\set{\sigma\in\sec[\gamma]{E}}{\text{$\sigma$ is $C^2$ with $\sigma(t_0)=0$ and $\sigma(t_1)=0$}}.
\]
Define also the vector space $F_a\subset T_a\AJ$ by $F_a=\Xi_a(\Sigma_\gamma)$, i.e.
\begin{align*}
F_a=\bigl\{v\in T_a\AJ\,|\, &\text{there exists $\sigma\in\sec[\gamma]{E}$ such that}\\ 
&\qquad\qquad
\text{$\sigma(t_0)=0$, $\sigma(t_1)=0$ and $v=\rho^1(\chi_E(\sigma,a,\dot{\sigma}))$}\bigr\}.
\end{align*}
and $F=\cup_{a\in\AJ} F_a\subset T\AJ$. 
\end{definition}

\begin{theorem}
The following properties hold.
\begin{enumerate}
\item For every $a\in\AJ$, the restriction of $\Xi_a$ to $\Sigma_\gamma$ is injective. Therefore, it provides an isomorphism between the real vector spaces $\Sigma_\gamma$ and $F_a$.
\item The codimension of $F$ is equal to $\dim(E)$.
\item $F$ is a smooth integrable subbundle of the tangent bundle to $\AJ$. 
\item The leaves of the foliation defined by $F$ are the $E$-homotopy classes.
\end{enumerate}
\end{theorem}
\begin{proof}
Item (1) follows directly from Corollary~\ref{Xi.injective}. Let $n=\dim(M)$ and $m=\rank(E)$ so that $\dim(E)=n+m$. For every $a\in\AJ$ the elements $Z$ of $F_a$ are determined by the $n+m$ independent equations 
\begin{itemize}
\item $T\tau(Z(t_0))=0$ ($n$ equations) which by Corollary~\ref{Xi.injective} implies that $Z=\Xi_a(\sigma)$ with $\sigma(t_0)=0$, and
\item $\sigma(t_1)=0$ ($m$ equations).
\end{itemize}
Thus the codimension of $F_a$ in $T_a\AJ$ is $n+m=\dim(E)$.

Denote by $\cals_J$ the set of time dependent sections $\eta$ of $E$ such that $\eta(t_0,-)=0$ and $\eta(t_1,-)=0$ and are $C^2$ in the variable $t$. For every section $\eta\in\cals_J$ we define the vector field $X_\eta$ on $\AJ$ by $X_\eta(a)=\Xi_a(\eta\circ\bar{\gamma})=\rho^1\circ\pr_2\circ\bar{\eta}\spC\circ\bar{a}$, where we recall that $\bar{\gamma}$ and $\bar{\eta}$ denote the time dependent objects corresponding to $\gamma$ and $\eta$. Then $X_\eta$ is tangent to $\AJ$ and it is clear that $X_\eta$ is a section of $F$. Moreover, every element of $F$ is of this form: if $v\in F_a$ for $a\in\AJ$, then  $v=\Xi_a(\sigma)$ for some curve $\sigma$ over $\gamma=\tau\circ a$ such that $\sigma(t_0)=\sigma(t_1)=0$. Let $\eta$ be any time-dependent section such that $\eta(t,\gamma(t))=\sigma(t)$ for all $t\in J$. Since $\sigma(t_0)=\sigma(t_1)=0$, we can take $\eta$ in $\cals_J$. Then the difference between $v$ and $X_\eta(a)$ is vertical (both project to $\rho(\sigma)$) and since $\sigma$ vanishes at $t_1$ it follows that they coincide.

This proves that the subbundle $F$ is spanned by vector fields of the form $X_\eta$ and hence $F$ is smooth. 

Given $\eta_1,\eta_2\in\cals_J$ we have that $[\eta_1,\eta_2]\in\cals_J$. Since the bracket $[\bar{\eta}\spC_1,\bar{\eta}\spC_2]$ of two complete lifts is the complete lift of the bracket $[\bar{\eta}_1,\bar{\eta}_2]\spC$, and $\rho^1$ and $\pr_2$ are morphisms of Lie algebroids, we have that $[X_{\eta_1},X_{\eta_2}]=X_{[\eta_1,\eta_2]}$. This proves that $F$ is involutive, and hence integrable.

Let $a_0\in\AJ$ and consider the homotopy class $\mathcal{H}$ of $a_0$. Consider also the integral leaf $\calf$ of the integrable subbundle $F$ which contains $a_0$. We will prove that both sets are equal, $\calf=\mathcal{H}$:
\begin{itemize}
\item[$\calf\subset\mathcal{H}$] Let $a_1\in\calf$. Then there exists a curve $s\mapsto a_s$ from $a_0$ to $a_1$ contained in $\calf$, that is such that $v_s=\frac{d}{ds}a_s\in F_{a_s}$. We can assume that $a_s$ is an integral curve of a vector field $X_\eta$ for some section $\eta\in\cals_J$ which moreover has compact support. Thus $a_s(t)=\pr_2(\Phi_s(\bar{a}_0(t)))$ defined for $(s,t)\in I\times J$. Then   $\xi=a(s,t)dt+\eta(t,\gamma(s,t))ds$, with $\gamma(s,t)=\tau(a(s,t))$, is a homotopy from $a_0$ to $a_1$, so that $a_1\in\mathcal{H}$.

\item[$\mathcal{H}\subset\calf$] Let $a_1\in\mathcal{H}$. Then there exists an $E$-homotopy $\Phi=adt+bds$ from $a_0$ to $a_1$. From the morphism condition we get $\pd{a}{s}(s,t)=\rho^1\chi_E(b,a,\pd{b}{t})$ which is an element of $F$. Thus the curve $a_s:t\mapsto a(s,t)$ is a curve in $\calf$ which ends at $a_1$. Hence $a_1\in\calf$.
\end{itemize}
This completes the proof. 
\end{proof}

We finally mention that a Lie algebroid $E$ is integrable, that is, $E$ is the Lie algebroid of some Lie groupoid, if and only if the foliation $F$ is a regular foliation and hence the set of $E$-homotopy classes, $\G=\AJ/\!\sim$, inherits a structure of smooth quotient manifold which makes it a smooth Lie groupoid over the manifold $M$. Moreover, it is the unique source simply-connected Lie groupoid with Lie algebroid $E$. See~\cite{Rui} for the details.

\section{The space of $E$-paths}
\label{PathSpace}
On the same set $\AJ$ there are two natural differential manifold structures: as a submanifold of the set of $C^1$ paths in $E$, which will be denoted just $\AJ$, and the structure induced by the foliation into $E$-homotopy classes, which will be denoted $\PJ$. The structure of $\AJ$ is relevant when one wants to study the relation between neighbor $E$-homotopy classes, as it is the case in the problem of integrability of Lie algebroids to Lie groupoids. We will show that the structure of $\PJ$ is just the structure that one needs in Mechanics, where one does not have the possibility to jump from one $E$-homotopy class to another. 

\subsection{Manifold structure}
The global version of Frobenius theorem and some of its consequences can be stated as follows.
\begin{theorem}[\cite{Lang,Michor}]
Let $F\subset TX$ be an integrable vector subbundle of $TX$. Using the restrictions of distinguished charts to plaques as charts we get a new structure of a smooth manifold on $X$, which we denote by $X_F$. If $F\not=TX$ the topology of $X_F$ is finer than that of $X$. $X_F$ has
uncountably many connected components, which are the leaves of the foliation, and the identity induces an injective immersion $\map{i}{X_F}{X}$. 

If $\map{f}{Y}{X}$ is a smooth map such that $Tf(TY)\subset F$, then the induced map $\map{f_F}{Y}{X_F}$ (same values $f_F(x)=f(x)$ but different differentiable structure on the target space) is also smooth. 
\qed
\end{theorem}

In our case, since the partition into $E$-homotopy classes defines a foliation on $\AJ$ it is natural to consider in the set $\AJ$ the structure of differentiable Banach manifold induced by such foliation, as explained above. We will denote this manifold by $\PJ$, that is $\PJ=\AJ_F$, and we will refer to it as the space of $E$-paths on the Lie algebroid $E$. Every homotopy class is a connected component of $\PJ$, and the identity defines a smooth map $\map{i}{\PJ}{\AJ}$ which is an (invertible) injective immersion. The image by $i$ of a leaf is an immersed (in general not embedded) submanifold of $\AJ$. The tangent space to $\PJ$ at $a$ is $T_a\PJ=F_a$. The topology of $\PJ$ is finer than the topology on $\AJ$. In particular, if $\map{G}{\AJ}{Y}$ is a smooth map, then $\map{G\circ i}{\PJ}{Y}$ is also smooth.

\subsection{Mappings induced by morphisms}

We recall that admissible maps are precisely those maps which transforms admissible curves into admissible curves. Therefore an admissible map $\map{\Phi}{E}{E'}$ induces a map between $E$-paths by composition $a\mapsto \Phi\circ a$. We prove now that such a map is smooth provided that $\Phi$ is a morphism.

\begin{proposition}
Given a morphism of Lie algebroids $\map{\Phi}{E}{E'}$ the induced map  $\map{\hat{\Phi}}{\PJ}{\PJ[E']}$ given by $\hat{\Phi}(a)=\Phi\circ a$ is smooth. 
\end{proposition}
\begin{proof}
We consider the auxiliary map $\map{\tilde{\Phi}}{\AJ}{\AJ[E']}$ given by $\tilde{\Phi}(a)=\Phi\circ a$ which is smooth. Then by composition with $\map{i}{\PJ}{\AJ}$, which is also smooth, we get a smooth map  $\bar{\Phi}=i\circ\tilde{\Phi}$ from $\PJ$ to $\AJ[E']$. We just need to prove (see~\cite{Lang}) that $T\bar{\Phi}$ maps $T\PJ$ into $F'$, the integrable subbundle in $\AJ[E']$. From proposition~\ref{morphism} we have that $T\Phi\circ\Xi_a(b)=\Xi_{\Phi\circ a}(\Phi\circ b)$. In other words, $T_a\bar{\Phi}(\Xi_a(b))=\Xi_{\Phi\circ a}(\Phi\circ b)$ which is an element of $F'_{\bar{\Phi}(a)}$.
\end{proof}

Many properties of $\hat{\Phi}$ are consequence of those of $\Phi$, as it is shown next. 

\begin{proposition}
Let $\map{\Phi}{E}{E'}$ be a morphism of Lie algebroids. 
\begin{itemize}
\item If $\Phi$ is fiberwise surjective then $\hat{\Phi}$ is a submersion. 
\item If $\Phi$ is fiberwise injective then $\hat{\Phi}$ is a immersion. 
\end{itemize}
\end{proposition}
\begin{proof} We will use that, for every $E$-path $a$, the map $\Xi_a$ is an isomorphism from $\Sigma_{\gamma}$ to $F_a$, where $\gamma=\tau\circ a$.

Assume that $\Phi$ is fiberwise surjective. From $T_a\hat{\Phi}(\Xi_a(\sigma)) =\Xi_{\Phi\circ a}(\Phi\circ \sigma)$ we immediately deduce that $T_a\hat{\Phi}$ is surjective. Therefore we just have to prove that the kernel splits. The kernel of $T_a\hat{\Phi}$ is 
\[
\ker T_a\hat{\Phi}=\set{\Xi_a(\sigma)}{\Phi\circ \sigma=0}.
\]
Consider a splitting $\map{\xi}{\Im\Phi}{E}$ of the exact sequence of vector bundles $0\to\ker\Phi\xrightarrow{i}E\xrightarrow{j}\Im\Phi\to0$ and define the vector space 
\[
F_2=\set{\Xi_a(\sigma)}{\exists c\in\sec[\gamma]{E}\text{ such that }\sigma=\xi\circ\Phi\circ c}.
\]
Then if $F_1=\ker T_a\hat{\Phi}$ we have that the map $\map{\alpha}{T_a\PJ}{F_1\times F_2}$ given by 
\[
\alpha(\Xi_a(c))=(\Xi_a(c-\xi\circ\Phi\circ c),\Xi_a(\xi\circ\Phi\circ c))
\]
is obviously an isomorphism, so that $\ker T_a\hat{\Phi}$ splits. Since the above holds for every $a\in\PJ$ we have that $\hat{\Phi}$ is a submersion.

Assume that $\Phi$ is fiberwise injective. Let $a$ be an $E$-path and denote by $a'$ the transformed path $a'=\Phi\circ a$. From $T\hat{\Phi}(\Xi_a(\sigma)) =\Xi_{a'}(\Phi\circ \sigma)$ we immediately deduce that $T_a\hat{\Phi}$ is injective. Therefore we just have to prove that the image splits. The image of $T_a\hat{\Phi}$ is 
\[
\Im T_a\hat{\Phi}=\set{\Xi_{a'}(\sigma)}{\exists\,c\in\sec[\gamma]{E}\text{ such that }\sigma=\Phi\circ c}.
\]
Consider a splitting $\map{\xi}{E'/\Im\Phi}{E'}$ of the exact sequence of vector bundles $0\to\Im\Phi\xrightarrow{i}E'\xrightarrow{j}E'/\Im\Phi\to0$ and define the vector space 
\[
F_2=\set{\Xi_{a'}(\sigma)}{\exists\,c\in\sec[\gamma]{E'/\Im\Phi}\text{ such that }\sigma=\xi\circ c}.
\]
Then if $F_1=\Im T_a\hat{\Phi}$ we have that the map $\map{\alpha}{T_{a'}\PJ[E']}{F_1\times F_2}$ given by 
\[
\alpha(\Xi_{a'}(c))=(\Xi_{a'}(c-\xi\circ j\circ c),\Xi_{a'}(\xi\circ j\circ c))
\]
is obviously an isomorphism, so that $\Im T_a\hat{\Phi}$ splits. Since the above holds for every $a\in\PJ$ we have that $\hat{\Phi}$ is an immersion.
\end{proof}

As a consequence if $\Phi$ is fiberwise bijective, then $\hat{\Phi}$ is a local diffeomorphism, that is maps diffeomorphically a neighborhood of a point in an $E$-homotopy class into a neighborhood of a point in an $E$-homotopy class.

\section{Variational description}

We consider a Lagrangian function $L$ on a Lie algebroid $E$. Lagrange's equations determine a dynamical system on $E$ introduced in~\cite{Weinstein} for the regular case, and in~\cite{LMLA} for the general case. We consider the question of whether this differential equations can be obtained from a variational principle, by imposing adequate boundary conditions.

\subsection{The case of integrable Lie algebroids}
As argued in~\cite{Weinstein}, in the case of an integrable Lie algebroid $E=\CMcal{L}(\G)$, the natural boundary conditions for a variational principle on $E$ are given by elements of the Lie groupoid $\G$.  
\begin{theorem}[\cite{Weinstein}]
Let $L$ be a regular Lagrangian on the Lie algebroid $E$ and let $g$ be an element of a Lie groupoid $\G$ whose Lie algebroid is $E$. The critical points of the functional $a\mapsto \int L(a(t))dt$ on the space of admissible paths whose development begins at $\s(g)$ and ends at $g$ are precisely those elements of that space which satisfy Lagrange's equations.
\end{theorem}

We recall that the development of an $E$-path $a$ is the curve $\map{g}{J}{\G}$ such that $TL_{g^{-1}(t)}\dot{g}(t)=a(t)$ for every $t\in J$ and $g(t_0)=\bepsilon(\tau(a(t_0)))$. In this expressions $L_g$ is the left translation and $\bepsilon$ is the unit map in the groupoid $\G$.

\smallskip

After the integrability results in~\cite{Rui} we can reformulate this result in a way that makes no (explicit) reference to the groupoid at all. Let us consider $\G$ as the source simply-connected groupoid integrating the Lie algebroid $E$. An element $g$ of $\G$ is but an $E$-homotopy class and `the space of admissible paths whose development begins at $\s(g)$ and ends at $g$' is but $g$ considered as a set of $E$-paths. Taking also into account the results of \cite{LMLA}, we can eliminate the condition of the regularity of the Lagrangian, and thus we can reformulate Weinstein's result as follows.

\begin{quote}
Let $L$ be a Lagrangian on an integrable Lie algebroid $E$ and let $g$ be an element of the source simply connected Lie groupoid $\G$ integrating $E$. The critical points of the functional $a\mapsto \int L(a(t))dt$ on the set of curves $a\in g$ are precisely those elements of that space which satisfy Lagrange's equations.
\end{quote}

But an $E$-homotopy class is a manifold whether the Lie algebroid is integrable or not. Therefore, a similar statement holds for the general case of a Lie algebroid integrable or not, as we are going to see.

\subsection{The general case}
With the manifold structure that we have previously defined on the space of $E$-paths, we can formulate the variational principle in a standard way. Let us fix two points $m_0,m_1\in M$ and consider the set $\PJ_{m_0}^{m_1}$ of those $E$-paths with fixed base endpoints equal to $m_0$ and $m_1$, that is 
\[
\PJ_{m_0}^{m_1}=\set{a\in\PJ}{\tau(a(t_0))=m_0\quand \tau(a(t_1))=m_1}.
\]
We remark that $\PJ_{m_0}^{m_1}$ is a Banach submanifold of $\PJ$, since it is a disjoint union of Banach submanifolds (the $E$-homotopy classes of curves with base path connecting such points). On the contrary, there is no guaranty that the analog set $\AJ_{m_0}^{m_1}$ is a manifold (see~\cite{PiTa}).

\begin{theorem}
\label{variational}
Let $L\in\cinfty{E}$ be a Lagrangian function on the Lie algebroid $E$ and fix two points $m_0, m_1\in M$. Consider the action functional $\map{S}{\PJ}{\Real}$ given by $S(a)=\int_{t_0}^{t_1} L(a(t))dt$. The critical points of $S$ on the Banach manifold $\PJ_{m_0}^{m_1}$ are precisely those elements of that space which satisfy Lagrange's equations.
\end{theorem}
\begin{proof}
The action functional $S$ is a smooth function on $\PJ_{m_0}^{m_1}$. The tangent space to such manifold at $a\in\PJ_{m_0}^{m_1}$ is $F_a$, i.e. the set of vector fields along $a$ of the form $\Xi_a(\sigma)$ for $\sigma\in\Sigma_\gamma$, i.e. $\sigma\in\sec[\gamma]{E}$ with $\sigma(t_0)=\sigma(t_1)=0$. Taking into account that $\Xi(fb)=f\Xi_a(\sigma)+\dot{f}\sigma\spV_a$, for every function $\map{f}{J}{\Real}$, and following the steps in~\cite{LAGGM} we get (here and in what follows $d$ is the usual (Frechet) differential of a function on a manifold)
\begin{align*}
0=\pai{d S(a)}{\Xi_a(fb)}
&=\int_{t_0}^{t_1}[f(t)\pai{dL}{\Xi_a(\sigma)}+\dot{f}\pai{dL}{\sigma\spV_a}]dt\\
&=\int_{t_0}^{t_1}f(t)\Bigl[\pai{dL}{\Xi_a(\sigma)}+\frac{d}{dt}\pai{\theta_L\circ a}{\sigma}\Bigr]dt
+f\pai{\theta_L\circ a}{\sigma}\Big|_{t_0}^{t_1}\\
&=\int_{t_0}^{t_1}f(t)\pai{\delta L(\dot{a}(t))}{\sigma(t)}dt,
\end{align*}
where we recall that $\delta L$ is given by $\pai{\delta L(\dot{a}(t))}{\sigma(t)} =\pai{dL}{\Xi_a(\sigma)}-\frac{d}{dt}\pai{\theta_L\circ a}{\sigma}$.
Since this holds for every function $f$ and every section $\sigma\in\Sigma_{\gamma}$ it follows that the critical points are determined by the equation $\delta L(\dot{a}(t))=0$, that is, by the Lagrange's equations.
\end{proof}

Alternatively, one can restrict the action to each connected component, that is, to each $E$-homotopy class with base endpoints $m_0$ and $m_1$. Every such homotopy class is a Banach manifold and the action $S$ is a smooth function on it. The rest of the proof is as above.

\subsection{Reduction}
The variational structure of the problem is not broken by reduction. On the contrary, reduction being a morphism of Lie algebroids, preserves such structure. We saw that morphisms transforms admissible variations into admissible variations, so that they induce a map between path spaces. Therefore, a morphism induces relations between critical points of functions defined on path spaces, in particular between the solution of Lagrange's equations.

Consider a morphism $\map{\Phi}{E}{E'}$ of Lie algebroids and the induced map between the spaces of paths $\map{\hat{\Phi}}{\PJ}{\PJ[E']}$. Consider a Lagrangian $L$ on $E$ and a Lagrangian $L'$ on $E'$ which are related\footnote{We may allow $L'\circ\Phi=L+\dot{f}$ for $f\in\cinfty{M}$ which makes $S'(\hat{\Phi}(a))=S(a)+c$ with $c=f(m_1)-f(m_0)$, constant. By redefining the Lagrangian $L$ to be $L+\dot{f}$ the equations of motion remain the same and we need to consider only the case $L'\circ\Phi=L$. More generally, we even may allow the addition of the linear function associated to a $d$-closed section of $E^*$.}
 by $\Phi$, that is, $L=L'\circ\Phi$. Then the associated action functionals $S$ on $\PJ$ and $S'$ on $\PJ[E']$ are related by $\hat{\Phi}$, that is  $S'\circ \hat{\Phi}=S$. Indeed,
\[
S'(\hat{\Phi}(a))=S'(\Phi\circ a)
=\int_{t_0}^{t_1}(L'\circ\Phi\circ a)(t)\,dt
=\int_{t_0}^{t_1}(L\circ a)(t)\,dt
=S(a).
\]

The following result is already in~\cite{Weinstein} but the proof is different.
\begin{theorem}[\cite{Weinstein}]
\label{reconstruction}
Let $\map{\Phi}{E}{E'}$ be a morphism of Lie algebroids. Consider a Lagrangian $L$ on $E$ and a Lagrangian $L'$ on $E'$ such that $L=L'\circ\Phi$. If $a$ is an $E$-path and $a'=\Phi\circ a$ is a solution of Lagrange's equations for $L'$ then $a$ itself is a solution of Lagrange's equations for $L$.
\end{theorem}
\begin{proof}
Since $S'\circ \hat{\Phi}=S$ we have that $\pai{dS'(\hat{\Phi}(a))}{T_a\hat{\Phi}(v)} =\pai{dS(a)}{v}$ for every $v\in T_a\PJ_{m_0}^{m_1}$. If $\hat{\Phi}(a)$ is a solution of Lagrange's equations for $L'$ then $dS'(\hat{\Phi}(a))=0$, from where it follows that $dS(a)=0$.
\end{proof}

From the above relations between the action functionals it readily follows a reduction theorem.

\begin{theorem}[Reduction]
\label{reduction}
Let $\map{\Phi}{E}{E'}$ be a fiberwise surjective morphism of Lie algebroids. Consider a Lagrangian $L$ on $E$ and a Lagrangian $L'$ on $E'$ such that $L=L'\circ\Phi$. If $a$ is a solution of Lagrange's equations for $L$ then $a'=\Phi\circ a$ is a solution of Lagrange's equations for $L'$.
\end{theorem}
\begin{proof}
Since $S'\circ \hat{\Phi}=S$ we have that $\pai{dS'(\hat{\Phi}(a))}{T_a\hat{\Phi}(v)} =\pai{dS(a)}{v}$ for every $v\in T_a\PJ_{m_0}^{m_1}$. If $\Phi$ is fiberwise surjective, then $\hat{\Phi}$ is a submersion, from where it follows that $\hat{\Phi}$ maps critical points of $S$ into critical points of $S'$, i.e. solutions of Lagrange's  equations for $L$ into solutions of Lagrange's equations for $L'$.
\end{proof}

We can reduce partially a system and then reduce it again. The result obviously coincides with the obtained by the total reduction.

\begin{theorem}[Reduction by stages]
\label{reduction.by.stages}
Let $\map{\Phi_1}{E}{E'}$ and $\map{\Phi_2}{E'}{E''}$ be fiberwise surjective morphisms of Lie algebroids. Let $L$, $L'$ and $L''$ be Lagrangian functions on $E$, $E'$ and $E''$, respectively, such that $L'\circ\Phi_1=L$ and $L''\circ\Phi_2=L'$. Then the result of reducing first by $\Phi_1$ and later by $\Phi_2$ coincides with the reduction by $\Phi=\Phi_2\circ\Phi_1$.
\end{theorem}
\begin{proof}
It is obvious since $\Phi=\Phi_2\circ\Phi_1$ is also a fiberwise surjective morphism of Lie algebroids.
\end{proof}

This result as stated here seems to be trivial, but a relevant case of application of this theorem is the case of reduction by a Lie group by first reducing by a closed normal subgroup and later by the residual quotient group~\cite{CeMaRa}. As we will see in the next subsection, Abelian Routh reduction can also be studied in the above framework.

Finally, we mention that the reconstruction procedure can be understood as follows. Consider a fiberwise surjective morphism $\map{\Phi}{E}{E'}$ and the associated reduction map $\map{\hat{\Phi}}{\PJ}{\PJ[E']}$. Given an $E'$-path $a'\in\PJ[E']$ solution of the dynamics defined by the Lagrangian $L'$, we look for an $E$-path $a\in\PJ$ solution of the dynamics for the Lagrangian $L=L'\circ\Phi$, such that $a'=\hat{\Phi}(a)$. For that, it is sufficient to find a map $\map{\xi}{\PJ[E']}{\PJ}$ such that $\hat{\Phi}\circ\xi=\id_{\PJ[E']}$. Indeed, given the $E'$-path $a'$ solution for the reduced Lagrangian $L'$, the curve $a=\xi(a')$ is an $E$-path and satisfy $\Phi\circ a=a'$. From theorem~\ref{reconstruction} we deduce that $a$ is a solution for the original Lagrangian. Of course one has to define a map $\xi$ and different maps define different $E$-paths $a$ for the same $E'$-path $a'$. Explicit constructions of such maps by using connections can be found in~\cite{IMS, OrRa}.

\subsection{Examples} We present here some examples where the reduction process indicated above can be applied.
\nobreak
\subsubsection*{Lie groups}
Consider a Lie group $G$ and its Lie algebra $\g$. The map $\map{\Phi}{TG}{\g}$ given by $\Phi(g,\dot{g})=g^{-1}\dot{g}$ is a morphism of Lie algebroids, which is fiberwise bijective. As a consequence if $L$ is a left-invariant Lagrangian function on $TG$ and $L'$ is the projected Lagrangian on the Lie algebra $\g$, that is $L(g,\dot{g})=L'(g^{-1}\dot{g})$, then every solution of Lagrange's equations for $L$ projects by $\Phi$ to a solution of  Lagrange's equations for $L'$. Moreover, since $\Phi$ is surjective every solution can be found in this way: if the projection $\xi(t)=g(t)^{-1}\dot{g}(t)$ of an admissible curve $(g(t),\dot{g}(t))$ is a solution of $L'$, then $(g(t),\dot{g}(t))$ is a solution for $L$. Thus, the Euler-Lagrange equations on the group reduce to the Euler-Poincaré equations on the Lie algebra.

Generalizing the above example we have the case of a Lie groupoid and its Lie algebroid.

\subsubsection*{Lie groupoids}
Consider a Lie groupoid $\G$ over $M$ with source $\s$ and target $\t$,  and with Lie algebroid $E$. Denote by $T^{\s}\G\to\G$ the kernel of $T\s$ with the structure of Lie algebroid as integrable subbundle of $T\G$. Then the map $\map{\Phi}{T^{\s}\G}{E}$ given by left translation to the identity, $\Phi(v_g)=TL_{g^{-1}}(v_g)$ is a morphism of Lie algebroids, which is moreover fiberwise surjective. As a consequence, if $L$ is a Lagrangian function on $E$ and  $\boldsymbol{L}$ is the associated left invariant Lagrangian on $T^{\s}\G$, then the solutions of Lagrange's equations for $\boldsymbol{L}$ project by $\Phi$ to solutions of the Lagrange's equations. Since $\Phi$ is moreover surjective, every solution can be found in this way.

This is the reduction process used in ~\cite{Weinstein} to prove the variational principle.

\subsubsection*{Group actions}
We consider a Lie group $G$ acting free and properly on a manifold $Q$, so that the quotient map $\map{\pi}{Q}{M}$ is a principal bundle. We consider the standard Lie algebroid structure on $E=TQ$ and the associated Atiyah algebroid $E'=TQ/G\to M$. The quotient map $\map{\Phi}{E}{E'}$, $\Phi(v)=[v]$ is a Lie algebroid morphism and it is fiberwise bijective. Every $G$-invariant Lagrangian on $TQ$ defines uniquely a Lagrangian $L'$ on $E'$ such that $L'\circ\Phi=L$. Therefore every solution of the $G$-invariant Lagrangian on $TQ$ projects to a solution of the reduced Lagrangian on $TQ/G$, and every solution on the reduced space can be obtained in this way. Thus, the Euler-Lagrange equations on the principal bundle reduce to the Lagrange-Poincaré equations on the Atiyah algebroid.

\subsubsection*{Semidirect products}
Let $G$ be a Lie group acting from the right on a manifold $M$. We consider the Lie algebroid $E=TG\times M\to G\times M$ where $M$ is a parameter manifold, that is, the anchor is $\rho(v_g,m)=(v_g,0_m)$ and the bracket is determined by the standard bracket of vector fields on $G$, i.e. of sections of $TG\to G$, with the coordinates in $M$ as parameters. Consider also the transformation Lie algebroid $E'=\g\times M\to M$, where $\rho(\xi,m)=\xi_M(m)$, ($\xi_M$ being the fundamental vector field associated to $\xi\in\g$) and the bracket is determined by the bracket in the Lie algebra~$\g$. The map $\Phi(v_g,m)=(g^{-1}v_g,mg)$ is a morphism of Lie algebroids over the action map $\varphi(g,m)=mg$, and it is fiberwise surjective.

Consider a Lagrangian  $L$ on $TG$ depending on the elements of $M$ as parameters. Assume that $L$ is not left invariant  but that it is invariant by the joint action $L(g^{-1}\dot{g},mg)=L(\dot{g},m)$. We consider the Lagrangian $L'$ on $E'$ by $L'(\xi,m)=L(\xi_G(e),m)$, so that $L'\circ\Phi=L$. Then the parametric variables $m$ adquieres dynamics due to the group action (we can understand this as the dynamics of the system as seen from a moving frame) and solutions of Lagrange's equations for $L$ are mapped by $\Phi$ to solutions of Lagrange's equations for $L'$. This situation occurs for the heavy top, which will be considered as an example in section~\ref{LagrangeMultipliers}.

Thus, the Euler-Lagrange equations on the group, with parameters, reduce to the Euler-Poisson-Poincaré equations on the Lie algebra, also known as the Euler-Poincaré equations with advected parameters~\cite{HoMaRa}. A similar construction can be done in the case of a principal bundle and an associated bundle.

\subsubsection*{Abelian Routh reduction} In the case of a group action, assume that the Lie group is abelian. For simplicity, assume that we have just one cyclic coordinate $\theta$ and denote by $q$ the other coordinates, so that $L=L(q,\dot{q},\dot{\theta})$. The Lagrangian $L$ on $TQ$ projects to a Lagrangian $L'$ on $TQ/G$ with the same coordinate expression. The solutions for $L$ obviously project to solutions for $L'$.

The momentum $\mu=\pd{L}{\dot{\theta}}(q,\dot{q},\dot{\theta})$ is conserved and, provided that $L$ is regular, we can find $\dot{\theta}=\Theta(q,\dot{q},\mu)$. The Routhian $R(q,\dot{q},\mu)=L(q,\dot{q},\Theta(q,\dot{q},\mu))-\mu\dot{\theta}$ when restricted to a level set of the momentum $\mu=c$ defines a function $L''$ on $T(Q/G)$ which is just $L(q,\dot{q})=R(q,\dot{q},c)$. Thus $L''(q,\dot{q})=L(q,\dot{q},\Theta(q,\dot{q},c))-\frac{d}{dt}(c\theta)$, i.e. $L$ and $L''$ differ on a total derivative.  Thus the actions for $L$ and $L''$ differ by a constant and Lagrange equations reduce to $T(Q/G)$.

Obviously the same construction can also be done for a general Abelian group of symmetry, but it does not generalize to the non-Abelian case. Notice that this is an example of reduction by stages; we first reduce from $TQ$ to $TQ/G$ and later we reduce from $TQ/G$ to $T(Q/G)$, thought of as a level set of the momentum.

\medskip

Let me finally mention that all this examples can also be studied in the context of the symplectic formalism on Lie algebroids, see~\cite{NHLSLA} or~\cite{SLMCLA}.

\section{Lagrange Multipliers}
\label{LagrangeMultipliers}
We can analyze the problem from the perspective of Lagrange multiplier method by imposing a condition on $\AJ$ which represents the constraint that our $E$-paths are in a given $E$-homotopy class. This is connected with the theory of \textsl{Lin constraints}~\cite{CeIbMa}. We consider only the case of an integrable Lie algebroid, since in the contrary we will not have a differential manifold structure in the set of $E$-homotopy equivalence classes.

Since there are many versions of what one calls Lagrange multiplier method, we will state clearly the one that we will use. Consider two Banach manifolds $U$ and $V$ and a differentiable map $\map{G}{U}{V}$. Assume that $G$ is a submersion, and for $c\in V$ we consider the submanifold $C=\set{u\in U}{G(u)=c}$. For a differentiable function $\map{F}{U}{\Real}$ we look for the critical points of $F$ subjected to the constraint $G(u)=c$, that is, the critical points of the restriction of $F$ to the submanifold $C$. 

\begin{theorem}[\cite{MTA}] The function $F$ has a critical point at $u_0\in C$ constrained by $G(u)=c$ if and only if there exists $\lambda\in T^*_cV$ such that 
$dF(u_0)=\lambda\circ T_{u_0}G$.\qed
\end{theorem}

\medskip

In the case on an integrable Lie algebroid $E$, the foliation defined by the $E$-homotopy equivalence relation is a regular foliation so that quotient $\G=\AJ/\sim$ has the structure of quotient manifold and the quotient projection $\map{q}{\AJ}{\G}$ is a submersion. Defining the source and target maps by $\s([a])=\tau(a(t_0))$ and $\t([a])=\tau(a(t_1))$, the unit map $\map{\bepsilon}{M}{\G}$ by $\bepsilon(m)=[0_m]$, where $0_m$ denotes the constant curve with value $0\in E_m$, and the multiplication induced by concatenation of $E$-paths, we have that $\G$ is the source simply-connected Lie groupoid with Lie algebroid $E$. See~\cite{Rui} for the details.

Given $g\in\G$, we can select the curves in an $E$-homotopy class as the set $q^{-1}(g)$. Therefore we look for the critical points of the functional $S(a)=\int_{t_0}^{t_1} L(a(t))\,dt$ defined in $\AJ$, constrained by the condition $q(a)=g$. Since $q$ is a submersion, there are not singular curves for the constraint map, and we can use Lagrange multiplier method in the version given above.

\begin{theorem}
Let $\map{S}{\AJ}{\Real}$, be the action functional $S(a)=\int_{t_0}^{t_1} L(a(t))\,dt$. An admissible curve $a\in\AJ$ is a solution of Lagrange's equations if and only if there exists $\mu\in T_g^*\G$ such that $dS(a)=\mu\circ T_aq$.
\end{theorem}

We may find more information about the value of the Lagrange multiplier and its relation to Lagrangian Mechanics by proceeding as follows. We first recall that the element $q(a)=g\in\G$ can be identified with the value at $t=t_1$ of the solution to the differential equation $TL_{\Gamma^{-1}}\dot{\Gamma}=a$ for a curve $\Gamma(t)$ in the $\s$-fiber $\s^{-1}(m_0)\subset \G$ with initial conditions $\Gamma(t_0)=\bepsilon(m_0)$. (In this expression and in what follows in this section $L_g$ denotes the left translation in the groupoid.) Thus we can consider $q$ as the endpoint mapping $q(a)=\Gamma(t_1)$ for the solution of such initial value problem. 

\begin{proposition}
The tangent map to $\map{q}{\AJ}{\G}$ at $a\in\AJ$ satisfies 
\[
T_aq(\Xi_a(\sigma))=T_{\bepsilon(m_1)}L_g(\sigma(t_1))
\]
where $\sigma\in\sec[\gamma]{E}$ is such that $\sigma(t_0)=0$.
\end{proposition}
\begin{proof}
Let $a_0\in\AJ$ be an $E$-path and consider the vector $\Xi_{a_0}(\sigma)$. Choose a time dependent section $\eta$ with compact support such that $\eta(t,\gamma(t))=\sigma(t)$ and the associated morphism $\xi=a(s,t)dt+b(s,t)ds$ from $T(I\times J)\subset T\Real^2$ to $E$, which satisfies $a(0,t)=a_0(t)$ and $b(s,t_0)=0$. Since $E$ is integrable, $\xi$ can be lifted to a morphism of Lie groupoids, $\map{\Lambda}{(I\times J)\times (I\times J)}{\G}$. This morphism being  defined on a pair groupoid, it is necessarily of the form $\Lambda(s,t,s',t')=\Gamma(s,t)^{-1}\Gamma(s',t')$ for some map $\map{\Gamma}{I\times J}{\G}$, and we can fix the value $\Gamma(0,t_0)=\bepsilon(m_0)$. Since the differential of $\Lambda$ restricted to the vectors tangent to the $\s$-fiber at the identity is to be equal to $\xi$ we have that 
\[
a(s,t)=TL_{\Gamma(s,t)^{-1}}\pd{\Gamma}{t}(s,t)
\qquand
b(s,t)=TL_{\Gamma(s,t)^{-1}}\pd{\Gamma}{s}(s,t).
\]
At $t=t_0$ we have that
$
0=b(s,t_0)=TL_{\Gamma(s,t_0)^{-1}}\pd{\Gamma}{s}(s,t_0)
$,
so that $\pd{\Gamma}{s}(s,t_0)=0$ and hence $\Gamma(s,t_0)$ is constant. Since at $s=0$ it evaluates to $\Gamma(0,t_0)=\bepsilon(m_0)$, we have that $t\mapsto \Gamma(s,t)$ is a solution of the initial value problem above for every $s$. We deduce that $q(a_s)=\Gamma(s,t_1)$, and hence 
\[
\der{}{s}\at{s=0}q(a_s)=\der{}{s}\Gamma(s,t_1)\at{s=0}=TL_{\Gamma(s,t_1)}b(s,t_1)\at{s=0}=TL_g\sigma(t_1).
\] 
which proves the result.
\end{proof}

In view of this result, if we apply Lagrange multiplier equation $\pai{dS(a)}{v}=\pai{\mu}{T_ap(v)}$ to the vector $v=\Xi_a(\sigma)$ and we integrate by parts as in the proof of theorem~\ref{variational} we get
\[
\int_{t_0}^{t_1}\pai{\delta L(\dot{a}(t))}{\sigma(t)}\,dt +\pai{\theta_L(a(t_1))}{\sigma(t_1)}=\pai{\mu}{T_{\bepsilon(m_1)}L_g(\sigma(t_1))}.
\]
For a solution $a$ of Lagrange equations, $\delta L(\dot{a}(t))=0$, and since $\sigma(t_1)\in E_{m_1}$ is arbitrary, we have that the multiplier satisfies 
$$
\pai{\theta_L(a(t_1))}{b}=\pai{\mu}{TL_g(b)}\qquad\text{for every $b\in E_{m_1}$.}
$$
Notice however that this equation determines the value of the multiplier $\mu$ only over vectors tangent to $\s^{-1}(m_1)$, the $\s$-fiber at $m_1$.

\medskip

Alternatively one can proceed as follows. Once we have fixed an element $g\in\G$ with source $\s(g)=m_0$ and target $\t(g)=m_1$, we consider the subset $\AJ_{m_0}$ of those $E$-paths whose base path start at $m_0$, that is 
\[
\AJ_{m_0}=\set{a\in\AJ}{\tau(a(t_0))=m_0}.
\]
\begin{proposition}
$\AJ_{m_0}$ is a submanifold of $\AJ$ and its tangent space at $a\in\AJ_{m_0}$ is 
\[
T_a\AJ_{m_0}=\set{\Xi_a(\sigma)}{\sigma(t_0)=0}.
\] 
\end{proposition}
\begin{proof}
Let $\map{\tilde{\s}}{\AJ}{M}$ be the map $\tilde{\s}(a)=\tau(a(t_0))$. We have that $T_a\tilde{\s}(Z)=T\tau\circ Z(t_0)$. For the uniqueness and existence theorem for initial value problems (applied to the `differential equation' $T_aG(Z)=0$, i.e. $Z\in T_a\AJ$) we have that given $w\in T_{m_0}M$ there exists $Z\in T_a\AJ$ such that $T\tau\circ Z(t_0)=w$. Therefore $T_a\tilde{\s}$ is surjective. 

By corollary~\ref{Xi.injective}, the kernel of $T_a\tilde{\s}$ is 
\[
\ker(T_a\tilde{s})=\set{\Xi_a(\sigma)}{\sigma(t_0)=0}.
\]
Then the kernel splits: if we take a subspace $C\subset T_{m_0}M$ complementing $\Im(\rho_{m_0})$ then the set of vector $Z\in T_a\AJ$  such that $T\tau\circ Z(t_0)\in C$ is a complementary subset of $\ker(T_a\tilde{s})$.
\end{proof}

On the submanifold $\AJ_{m_0}$ we define the map $\map{p}{\AJ_{m_0}}{\s^{-1}(m_1)}$ by $p(a)=L_{g^{-1}}(q(a))$. With the help of this map, the constraint reads $p(a)=\bepsilon(m_1)$, because an $E$-path is in $q^{-1}(g)$ if and only if it is in  $p^{-1}(\bepsilon(m_1))$. Then $p$ is a submersion and the tangent map $\map{T_ap}{T_a\AJ_{m_0}}{E_{m_1}}$ to $p$ at $a\in p^{-1}(\bepsilon(m_1))$, satisfies
\[
T_ap(\Xi_a(\sigma))=\sigma(t_1)
\]
for every $\sigma\in\sec[\gamma]{E}$ such that $\sigma(t_0)=0$. If we now apply Lagrange multiplier theorem we obtain the following result.

\begin{theorem}
Let $S_{m_0}$ be the restriction of the action functional to the submanifold $\AJ_{m_0}$. An admissible curve $a\in\AJ_{m_0}$ is a solution of Lagrange's equations if and only if it there exists $\lambda\in E^*_{m_1}$ such that  $dS_{m_0}(a)=\lambda\circ T_aq$. The multiplier $\lambda$ is given explicitly by $\lambda=\theta_L(a(t_1))$.
\end{theorem}
\begin{proof}
The constraint map is a submersion, so that $a\in\AJ_{m_0}$ is a constrained critical point if and only if there exists $\lambda\in T^*_{\bepsilon(m_1)}(\s^{-1}(m_1))=E^*_{m_1}$ such that $dS_{m_0}(a)=\lambda\circ T_ap$. We have to prove that the multiplier $\lambda$ is given by the momenta at the endpoint, $\lambda=\theta_L(a(t_1))$. If we apply Lagrange multiplier equation $\pai{dS(a)}{v}=\pai{\lambda}{T_ap(v)}$ to the vector $v=\Xi_a(\sigma)$ and we integrate by parts as in the proof of theorem~\ref{variational} we get
\[
\int_{t_0}^{t_1}\pai{\delta L(\dot{a}(t))}{\sigma(t)}\,dt =\pai{\lambda-\theta_L(a(t_1))}{\sigma(t_1)}.
\]
For a solution $a$ of Lagrange equations, $\delta L(\dot{a}(t))=0$, and since $\sigma(t_1)$ is arbitrary, we have that the multiplier is given by $\lambda=\theta_L(a(t_1))$.
\end{proof}

\subsubsection*{The heuristic Lagrange multiplier method}
One should notice that the above arguments by no means imply that we can use Lagrange multipliers rule in the `finite dimensional' form to which we refer as the `heuristic' method. That is, if we look for critical points of the action with the constraints $\dot{x}^i=\rho^i_\alpha y^\alpha$ then we can consider the extended Lagrangian on $\mathcal{L}\in\cinfty{TM\oplus T^*M\oplus E}$ given by $\mathcal{L}=p_0L+p_i(\dot{x}^i-\rho^i_\alpha y^\alpha)$, with $p_0\in\Real$, and we find the Euler-Lagrange equations for $\mathcal{L}$, 
\[
\pd{\mathcal{L}}{y^\alpha}=0
\qquad\qquad
\pd{\mathcal{L}}{p_i}=0
\qquand
\der{}{t}\pd{\mathcal{L}}{\dot{x}^i}-\pd{\mathcal{L}}{x^i}=0,
\]
then we usually find only a subset of the solution set. For normal solutions, i.e. with $p_0=1$, the equations we find are 
\[
p_i\rho^i_\alpha=\pd{L}{y^\alpha},
\qquad\qquad
\dot{x}^i=\rho^i_\alpha y^\alpha
\quad\qquand\quad
\dot{p_i}=\pd{L}{x^i}-\pd{\rho^j_\alpha}{x^i}y^\alpha.
\]
After some straightforward manipulations, taking the total derivative of the first equation, using the third equation and the structure equations, we get the Euler-Lagrange equations
\[
\dot{x}^i=\rho^i_\alpha y^\alpha  
\qquad\qquad
\frac{d}{dt}\left(\pd{L}{y^\alpha}\right)
=\rho^i_\alpha\pd{L}{x^i}-C^\gamma_{\alpha\beta}y^\beta\pd{L}{y^\gamma},
\]
but we should not forget the first equation $p_i\rho^i_\alpha=\pd{L}{y^\alpha}$, which imposes very severe restrictions to the value of the momenta $\pd{L}{y^\alpha}$ along solutions. For instance the momenta must vanish when contracted with elements of the kernel of the anchor. Abnormal solutions, i.e. for $p_0=0$ with $p\neq0$, can also exists, and are those admissible curves for which there exists $p_i$ such that $p_i\rho^i_\alpha=0$ and are solution of the differential equation $\dot{p_i}+p_j\pd{\rho^j_\alpha}{x^i}y^\alpha=0$. In many cases all admissible curves are abnormal solutions, so that all admissible curves are candidates and the method gives no information at all. In other cases, there are no abnormal solutions, but the variational equations predict the existence of some of them. In the next subsection we will show a physical example where both situations are shown explicitly.

On the other hand, the equations obtained by Lagrange multiplier trick can also be obtained by applying Pontryagin maximum principle to our system where the admissibility constraints are considered as the control equations, the coordinates $y^\alpha$ being the controls. In this respect, we mention that a way to do reduction, in the spirit of the results in this paper, and in the context of optimal control theory, was stated in~\cite{ROCT}.

\subsubsection*{An example: Lagrange top}
As a concrete example we consider a symmetric heavy top in body coordinates. This is a particular case of what we called a system on a semidirect product. The Lagrangian  is
$$
L(\gamma,\omega)=\frac{1}{2}\omega\cdot I\omega -\gamma\cdot e,
$$ 
where $\gamma\in\Real^3$ is (proportional to) the gravity direction and $\omega\in\mathfrak{so}(3)\simeq\Real^3$ is the body angular velocity both in the body reference frame, and $e$ is the unit vector in the direction of the symmetry axis of the top (and hence is constant). The gravity vector is constant in the space frame so that in the body frame it satisfies the constraint 
$
\dot{\gamma}=\gamma\times\omega
$.
Our configuration space is a Lie algebroid: the transformation Lie algebroid $\map{\tau=\pr_1}{\Real^3\times\mathfrak{so}(3)}{\Real^3}$ associated to the standard action of the Lie algebra of the rotation group on $\Real^3$. The above constraint is but the admissibility condition for a curve on $E$.

Admissible variations are of the form
\begin{equation}
\Xi_{(\gamma,\omega)}(\sigma)=(\gamma\times\sigma)\pd{}{\gamma}
 +(\dot{\sigma}+\omega\times\sigma)\pd{}{\omega},
\end{equation}
for every function $\map{\sigma}{\Real}{\Real^3}$, vanishing at $t_0$ and $t_1$, or in more traditional notation,
\[
\delta\gamma = \gamma\times\sigma
\qquand
\delta\omega = \dot{\sigma}+\omega\times\sigma.
\]
From here we get that Lagrange's equations are
\begin{equation}\label{Euler-top}
\begin{gathered}
I\dot{\omega}+\omega\times(I\omega)=\gamma\times e\\
\dot{\gamma}=\gamma\times\omega.
\end{gathered}
\end{equation}
in agreement with the classical equations in the Newtonian mechanics. 

We apply now the heuristic Lagrange multiplier trick. The extended Lagrangian is 
$$
\mathcal{L}=p_0\left(\frac{1}{2}\omega\cdot I\omega -\gamma\cdot e\right) + p\cdot (\dot{\gamma}-\gamma\times\omega),
$$
where $p\in\Real^3$ is the vector of Lagrange multipliers and $p_0\in\Real$. 
Abnormal solutions are obtained by setting $p_0=0$, with the condition $p\neq0$, and we get the equations
\begin{equation}\label{Abnormal}
0=p\times\gamma
\qquad
\dot{\gamma}=\gamma\times\omega
\qquand
\dot{p}=-\omega\times p.
\end{equation}
It follows that any admissible curve and a real number $\alpha\neq0$ determine an abnormal solution by taking $p=\alpha\gamma$, and hence, the set of candidates to solution of our problem is precisely the whole set of admissible curves, and no information is given by the method. This is a clear counter-example to the common belief~\cite{Carlos} that in optimization problems it is simpler to add the abnormal curves into the family of possible optimals, and to find the global optimum among them.  

\medskip

The situation is even worse. It is clear (either from the proper nature of the problem or from the constraint equations) that the length of the gravity vector $\gamma$ is constant. Therefore, we can consider our system with the same analytical expression but with $\gamma$ in a sphere $S^2$. Thus our configuration space is the transitive Lie algebroid $E'=S^2\times\Real^3\to S^2$ (the restriction to a leaf of the initial Lie algebroid) and it is obvious that we again get the same equations~\eqref{Euler-top}. But now, when applying the Lagrange multiplier method, the multiplier $p\in T_\gamma S^2$ is orthogonal to $\gamma$ and there are not abnormal solutions. On the other hand, normal solutions, satisfy $p\times\gamma=I\omega$, so that there are solutions of~\eqref{Euler-top} which are not obtained by using the Lagrange multiplier trick. For instance, for a symmetric top, $e$ is an eigenvector of $I$ and the relative equilibria solution $\gamma=\pm e$, $\omega=e$ (upward/downward spinning top) does not satisfy the restriction $p\times\gamma=I\omega$.

\section{Conclusions}
{\parskip=3pt plus 1pt 
We have shown that the Euler-Lagrange equations for a Lagrangian system on a Lie algebroid are the equations for the critical points of the action functional defined on the space of $E$-paths on the Lie algebroid. It should again be stressed that this variational principle is a `true' variational principle, that is, variations are curves in a manifold of curves satisfying the admissibility constraints and the action is stationary for every such variation in that manifold. It is not a variational principle of nonholonomic\footnote{In nonholonomic mechanics different Lagrangians which coincides when restricted to the constraint manifold give different Lagrange-D'Alembert equations.} or Hölder type (where only a subset of infinitesimal variations are considered which moreover are not tangent to the constraint manifold), it is neither a vakonomic\footnote{We recall that \textsc{vak}onomic means of \textsc{v}ariational \textsc{a}xiomatic \textsc{k}ind (see~\cite{DynamicalSystemsIII}).} principle (where the variations are assumed to satisfy the constraints only infinitesimally, and was introduced in order to solve some rigidity problems~\cite{DynamicalSystemsIII}, being therefore an \textit{axiomatic} way to define the dynamics).

We have also shown that reduction by a symmetry group, or by other more general fiberwise surjective morphisms of Lie algebroids, does not destroy the variational character of the problem. If we have a variational problem on a Lie algebroid, then the reduced problem is also variational, and the given morphism maps admissible variations for the original problem into  admissible variations for the reduced one. 

From our results it is clear the equivalence (in the appropriate particular cases) of our variational principle with some results stated in the literature under the name Euler-Poincaré or Lagrange-Poincaré variational principles, and which allows to obtain the so called Euler-Poincaré or Lagrange-Poincaré equations for systems with symmetry. Strictly speaking, Lagrange-Poincaré equations~\cite{CeMaRa} are the intrinsec expression of the equation of motion obtained by using the additional geometric structures carried by Lagrange-Poincaré bundles. See~\cite{CeMaPeRa} for a review, and see also~\cite{Carlos} for a recompilation of such results, where the non variationality of such equations is stated without proof. It would be interesting to study Hamilton-Poincaré variational principles~\cite{CeMaPeRa} in this setting.

It follows that one can obtain Lagrange's equations (and hence Euler-Poincaré, Lagrange-Poincaré, etc) either by a standard variational principle (critical points of a smooth function on an adequate Banach manifold), or by a `generalized variational principle' by considering as infinitesimal variations only those associated to complete lifts. This will be helpful when considering the second variation.

We have shown that, at least in the integrable case, one can also obtain such equations by using Lagrange multiplier method. This is connected to the results in~\cite{CeIbMa} where some restrictions, known as Lin constraints, need to be imposed to the variational problem in order to get the right equations of motion. 

I have also shown in an example that  what I called the `heuristic' Lagrange multiplier method, cannot be used to obtain Euler-Lagrange equations on Lie algebroids. There is nothing wrong with this method as long as one recognizes that it is a heuristic method, that is, there is no warranty that the candidates predicted by the method are solutions neither that all solutions appear as a candidate. It is in general a good help to infer the correct equations~\cite{LEALA}. In particular, our results shows that the method can perfectly work before reduction while it may not work after reduction. See~\cite{Carlos}, where such method is proposed as a key ingredient in the calculus of variations and an unavoidable part of the process.

I would like to stress the fact that the variational character of the equations of motion has nothing to do with the integrability of the Lie algebroid by a Lie groupoid. The integrability problem is related to the differentiable structure of the quotient $\AJ/\sim$, while for the variational description we only used the structure of $\PJ/\sim$, which is discrete over $M\times M$. In my opinion, it has to do with the `integrability' condition imposed by the Jacobi identity, which is necessary to prove that complete lifts form a Lie subalgebra. In this respect, it would be interesting to see if the results of this paper can be extended to the more general case of an anchored bundle or a general algebroid~\cite{GrGrUr}. 

Connected with the above ideas, let us finally mention that our arguments correspond essentially to the following idea, which was implicitly used in~\cite{LAGGM, CFTLAVA}. Let $\Sigma$ be an infinite dimensional Lie algebra of sections of a bundle acting on a manifold $P$, that is, there exists a morphism of Lie algebras $\map{\Theta}{\Sigma}{\vf{P}}$. Let $\map{F}{P}{\Real}$ a smooth function and consider $S:a\mapsto\int F(a(t))dt$, defined on the set $\mathcal{P}$ of curves which starts at $t_0$ on $p\subset P$ and ends at $t_1$ at $q\subset P$. If the function $S$ is stationary for every variation of the form $\Theta(\sigma)$ for $\sigma$ a time dependent section tangent to $p$ at $t_0$ and tangent to $q$ at $t_1$, then we can formulate a variational principle by restricting $S$ to the orbits of the induced action of the Lie algebra $\Sigma$ on the space of curves starting on $p$ and ending on $q$. Such action is $\sigma\in\Sigma\mapsto\tilde{\sigma}\in\vf{\mathcal{P}}$ with $\tilde{\sigma}(a)=\Theta(\sigma)\circ a$. In our case the Lie algebra of variations is the Lie subalgebra of complete lifts of sections of the Lie algebroid. While in general this can be considered as a somehow tautological procedure, in our case the foliation is intimately related to the geometry of the problem, and can be determined or reinterpreted in terms of such geometry.
}%end parskip parindent
%----------------------------------------------------------------------


\begin{thebibliography}{99}
  \let\\, 
  \newcommand{\by}[1]{\textsc{\ignorespaces #1}\\}
  \newcommand{\title}[1]{\textsl{\ignorespaces #1}\\}
  \newcommand{\vol}[1]{\textbf{\ignorespaces #1}}
  \newcommand{\info}[1]{\textrm{\ignorespaces #1}.}
  

\bibitem{MTA} 
  \by{Abraham R, Marsden JE and Ratiu TS}
  \title{Manifolds, tensor analysis and applications} 
  \info{Addison-Wesley, 1983}

\bibitem{Altafini}
\by{Altafini C}
   \title{Reduction by group symmetry of second order variational problems
    on a semidirect product of Lie groups with positive definite Riemannian
    metric}
    \info{\textsc{esaim:} Control, Optimisation and Calculus of Variations,
          \vol{10} (2004) 526--548}

\bibitem{DynamicalSystemsIII}
  \by{Arnold VI} 
  \title{Dynamical Systems III}
  \info{Springer-Verlag, 1988}

\bibitem{CannasWeinstein}
  \by{Cannas da Silva A and Weinstein A} 
  \title{Geometric models for noncommutative algebras} 
  \info{Amer. Math. Soc., Providence, RI, 1999; xiv+184 pp}

\bibitem{LAGGM}
   \by{Cari\~nena JF and Martínez E}
   \title{Lie algebroid generalization of geometric mechanics}
   \info{In Lie Algebroids and related topics in differential geometry (Warsaw 2000). Banach Center Publications 54 2001, p. 201}

\bibitem{CeIbMa}
  \by{Cendra H, Ibort A and Marsden JE}
  \title{Variational principal fiber bundles: a geometric theory of Clebsch potentials and Lin constraints}
  \info{J. Geom. Phys. \vol{4} (1987) 183--206}

\bibitem{CeMaRa} 
  \by{Cendra H, Marsden JE and Ratiu TS}
  \title{Lagrangian reduction by stages} 
  \info{Mem. Amer. Math. Soc. \vol{152} (2001), no.~722, x+108 pp}
  
\bibitem{CeMaPeRa}
  \by{Cendra H, Marsden JE, Pekarsky S and Ratiu TS}
  \title{Variational principles for Lie-Poisson and 
   Hamilton-Poincaré equations}
  \info{Moscow Mathematical Journal \vol{3}, (2003), 833--867}

\bibitem{NHLSLA} 
  \by{Cort\'es J, de Le\'on M, Marrero JC and Mart{\'\i}nez E} 
  \title{Nonholonomic Lagrangian systems on Lie algebroids}
  \info{Preprint 2005, arXiv:math-ph/0512003}

\bibitem{SLMCLA}
  \by{Cort\'es J, de Le\'on M, Marrero JC, Mart\'{\i}n de Diego D 
  and Mart\'{\i}nez E}
  \title{A survey of Lagrangian mechanics and control on Lie algebroids 
  and groupoids}
  \info{Int. Jour. on Geom. Meth. in Math. Phys. \vol{3} (2006) 509-558}

\bibitem{Rui}
  \by{Crainic M and Fernandes RL}
  \title{Integrability of Lie brackets}
  \info{Ann. of Math. (2) \vol{157} (2003), no. 2, 575--620}
  
\bibitem{Cr} 
  \by{Crampin M} 
  \title{Tangent bundle geometry for Lagrangian dynamics} 
  \info{J. Phys. A: Math. Gen. \vol{16} (1983) 3755--3772}

\bibitem{GrGrUr}
  \by{Grabowska K, Grabowski J and  Urbanski P}
  \title{Geometrical Mechanics on algebroids}
  \info{Int. Jour. on Geom. Meth. in Math. Phys. \vol{3} (2006) 559-576}

\bibitem{HoMaRa}
  \by{Holm DD, Marsden JE and Ratiu TS}
  \title{The Euler-Poincaré Equations and Semidirect Products with
         Applications to Continuum Theories} 
  \info{Adv. in Math. \vol{137} (1998) 1--81}

\bibitem{Klein} 
  \by{Klein J} 
  \title{Espaces variationnels et m\'ecanique} 
  \info{Ann. Inst. Fourier \vol{12} (1962) 1--124}

\bibitem{Lang}
  \by{Lang S}
  \title{Differential manifolds}
  \info{Springer-Verlag, New-York, 1972}

\bibitem{LSDLA}
  \by{de Le\'on M, Marrero JC and Mart\'{\i}nez E}
  \title{Lagrangian submanifolds and dynamics on Lie algebroids}
  \info{J. Phys. A: Math. Gen. \vol{38} (2005), R241--R308}

\bibitem{Carlos}
  \by{L\'opez C} 
  \title{Variational calculus, symmetries and reduction}
  \info{Int. Jour. on Geom. Meth. in Math. Phys. \vol{3} (2006) 577-590}

\bibitem{Mackenzie2}
  \by{Mackenzie KCH}
  \title{General Theory of Lie Groupoids and Lie Algebroids}
  \info{Cambridge University Press, 2005}
  %London Mathematical Society Lecture Note Series: 213, 

\bibitem{IMS}
  \by{Marsden JE and Ratiu TS} 
  \title{Introduction to Mechanics and symmetry}
  \info{Springer-Verlag, 1999}

\bibitem{LMLA}
  \by{Mart\'{\i}nez E}
  \title{Lagrangian Mechanics on Lie algebroids}
  \info{Acta Appl. Math., \vol{67} (2001), 295-320}
  
\bibitem{Medina} 
  \by{Mart\'{\i}nez E}
  \title{Geometric formulation of Mechanics on Lie algebroids}
  \info{In Proceedings of the VIII Fall Workshop on Geometry and Physics,
Medina del Campo, 1999, {\sl Publicaciones de la RSME}, \vol{2}
(2001), 209--222}

\bibitem{ROCT}
  \by{Mart\'{\i}nez E}
  \title{Reduction in optimal control theory}
  \info{Rep. Math. Phys. \vol{53} (2004) 79--90}

\bibitem{CFTLAVA}
  \by{Mart\'{\i}nez E}
  \title{Classical Field Theory on Lie algebroids: Variational aspects}
  \info{J. Phys. A: Mat. Gen. \vol{38} (2005) 7145-7160}

\bibitem{CFTLAMF}
  \by{Mart\'{\i}nez E}
  \title{Classical field theory on Lie algebroids: Multisymplectic formalism}
  \info{Preprint 2004, arXiv:math.DG/0411352}

\bibitem{LASLSAB}
  \by{Mart\'{\i}nez E, Mestdag T and Sarlet W}
  \title{Lie algebroid structures and Lagrangian systems on affine bundles}
  \info{J. Geom. Phys. \vol{44} (2002), no.~1, 70-95}

\bibitem{Michor}
  \by{Michor P}
  \title{Topics in differential geometry}
  \info{Book on the internet, unpublished}

\bibitem{OrRa}
  \by{Ortega JP and Ratiu TS}
  \title{Momentum maps and Hamiltonian Reduction}
  \info{Birkh\"auser, 2004}

\bibitem{PiTa}
  \by{Piccione P and Tausk D} 
  \title{Lagrangian and Hamiltonian formalism for constrained variational problems} 
  \info{Proc. Roy. Soc. Edinburgh Sect. A \vol{132} (2002) 1417--1437}
  
\bibitem{LEALA} 
  \by{Sarlet W, Mestdag T and Mart\'{\i}nez E}
  \title{Lagrangian equations on affine Lie algebroids} \info{
    Differential Geometry and its Applications, Proc.\ 8th Int.\
    Conf.\ (Opava 2001), D.\ Krupka et al, Editors}

\bibitem{Weinstein}
  \by{Weinstein A}
  \title{Lagrangian Mechanics and groupoids}
  \info{Fields Inst. Comm. \vol{7} (1996), 207-231}

\end{thebibliography}
\end{document}